\documentclass[aps,prl,reprint,preprintnumbers,amsmath,amssymb,floatfix,superscriptaddress,longbibliography]{revtex4-1}
\usepackage{float}
\usepackage{amssymb}
\usepackage{graphicx}
\usepackage{epsfig}
\usepackage{subcaption} 
\usepackage{dcolumn}
\usepackage{bbm}
\usepackage[colorlinks,linkcolor=red,anchorcolor=green,citecolor=blue,CJKbookmarks=True]{hyperref}
\usepackage{braket}
\usepackage{caption}
\usepackage{ragged2e}
\usepackage{multirow} 
\usepackage{tikz}
\usetikzlibrary{decorations.pathreplacing, positioning, arrows.meta}
\newcommand{\red}[1]{\textcolor{red}{#1}}

\begin{document}
\title{Portraits of Charmoniumlike States}

\author{\small Qingyang Liu}
\email{qyliu@ihep.ac.cn}
\affiliation{Institute of High Energy Physics, Chinese Academy of Sciences, Beijing 100049, People's Republic of China}
\affiliation{School of Physics, Huazhong University of Science and Technology, Wuhan 430074, People's Republic of China}

\author{\small Xiangyu Jiang}
\email{jiangxiangyu@ihep.ac.cn}
\affiliation{Institute of High Energy Physics, Chinese Academy of Sciences, Beijing 100049, People's Republic of China}
\affiliation{School of Physical Sciences, University of Chinese Academy of Sciences, Beijing 100049, People's Republic of China}

\author{\small Ying Chen}
\email{cheny@ihep.ac.cn}
\affiliation{Institute of High Energy Physics, Chinese Academy of Sciences, Beijing 100049, People's Republic of China}
\affiliation{School of Physical Sciences, University of Chinese Academy of Sciences, Beijing 100049, People's Republic of China}
\affiliation{Center for High Energy Physics, Henan Academy of Sciences, Zhengzhou 450046, People's Republic of China}

\author{\small Chunjiang Shi}
\affiliation{Institute of High Energy Physics, Chinese Academy of Sciences, Beijing 100049, People's Republic of China}
\affiliation{School of Physical Sciences, University of Chinese Academy of Sciences, Beijing 100049, People's Republic of China}

\author{\small Wei Sun}
\affiliation{Institute of High Energy Physics, Chinese Academy of Sciences, Beijing 100049, People's Republic of China}

\def\modified#1{\red{#1}}
\begin{abstract}
    We carry out the first lattice calculation of the charm quark density-density correlations for $1S$ and $1P$ conventional charmonia, as well as charmoniumlike states $\eta_{c1}(1^{-+})$ and $h_{c0}(0^{+-})$,  provide {\it ab initio} information of the charm quark distributions within these states. The spatial structures of $1S$ and $1P$ exhibit Clear relativistic effects that can be understood neatly in the Dirac theory of quarks where quark spin is naturally incorporated. For the charmoniumlike hybrid $\eta_{c1}(1^{-+})$, the relative distribution of $c\bar{c}$ indicates that its $c\bar{c}$ component has quantum numbers $1^{-(-)}$, such that $\eta_{c1}$ can be view as a bound state of a color octet $c\bar{c}$ and a $1^{+-}$ gluonic (chromomagnetic) componet in $S$-wave. This spatial configuration of $\eta_{c1}$ does not support the flux tube picture in the Born-Oppenheimer approximation.

\end{abstract}
\maketitle
\paragraph{Introduction---}
The existence of the charm quark predicted by Glashow-Iliopoulos-Maiani mechanism~\cite{Glashow:1970gm} was confirmed by the discovery of the $J/\psi$ particle in 1974~\cite{E598:1974sol,SLAC-SP-017:1974ind}, which is one of the bound states of charm quark and antiquark ($c\bar{c}$), namely, charmonium states or charmonia~\cite{Appelquist:1974yr,Eichten:1974af}. In the past fifty years, charmonia are mostly described by nonrelativistic quark models (NRQM)~\cite{Godfrey:1985xj} with various confining potentials and relativistic corrections~\cite{Voloshin:2007dx}. By solving the related Schr\"{o}dinger equation, NRQMs predict a populous family of charmonium states, whose members are usually labelled by $n^{2S+1}L_J$ with $n$, $S$, $L$, $J$ are the radial quantum number, the total spin of the $c\bar{c}$, the angular momentum, and the spin of the state, respectively. The predicted mass spectrum of lowest-lying states almost reproduces the experimental results, and the wave functions give descriptions of the internal structures of charmonia and are usually used to explain their electromagnetic transitions, leptonic decays and other properties. Besides charmonia, there may exist charmoniumlike states of a configuration $c\bar{c}g$, called charmoniumlike hybrids, where $g$ refers to gluonic degrees of freedom. It is intriguing to probe the structures of charmonia and charmoiniumlike hybrids through a model independent way, which provide more reliable information to the underlying dynamics of hadron formation. 

The internal structure of charmonium(like) states can be accessed by lattice QCD. One type of the observables is the equal-time Bethe-Salpeter wave functions~\cite{Velikson:1984qw,Barad:1984px,Hecht:1992uq,Hecht:1992ps,Babich:2007ah,Broniowski:2009dt} either in a fixed gauge~\cite{Chu:1990ps,Lissia:1991gv}. An alternative observable is the density-density correlation in a charmonium state $|H\rangle$~\cite{Barad:1984px,Wilcox:1986ge,Wilcox:1986dk,Wilcox:1990zc,Chu:1990ps,Lissia:1991gv,Burkardt:1994pw,Alexandrou:2002nn,Alexandrou:2003qt,Alexandrou:2008ru,Blossier:2016vgh}, namely,
$C_H(\vec{r})=\langle H|\rho(\vec{0})\rho(\vec{r})|H\rangle$, where $\rho(\vec{x})=[:\bar{c}\gamma_4 c:](\vec{x})$ is the charge density of charm quark in normal ordering. If we focus on the spatial correlation of the charm quark and charm antiquark within $|H\rangle$ and neglect the $c\bar{c}$ annihilation effects, we can view the charmonium state $|H\rangle$ as a bound state of a pair of quark and antiquark with different flavors, namely $c$ quark and $c'$ quark~\cite{Bali:2018nde}. In this sense, we replace $\rho(\vec{0})$ by $[:\bar{c}'\gamma_4 c':](\vec{0})$. After inserting the unit operator $\int d^3 x|\vec{x}\rangle\langle \vec{x}|=1$ between $\rho(\vec{0})$ and $\rho(\vec{r})$, we have~\cite{Wilcox:1990zc}
\begin{eqnarray}\label{eq:correlation}
    C_H(\vec{r})&\equiv&\int d^3 x \langle H|\rho(\vec{0})|\vec{x}\rangle\langle x|\rho(\vec{r})|H\rangle\nonumber\\
    &=&\int d^3 x |\Phi_{c'}(\vec{x})|^2|\Phi_{c}(\vec{x}+\vec{r})|^2,
\end{eqnarray}
where $|\Phi_{c{(')}}(\vec{x})|^2=\langle x|\rho(\vec{0})|H\rangle$ is the spatial charge density of the charm quark $c{(')}$ within $|H\rangle$ (in its rest frame). Since the state $|H\rangle$ distributes uniformly in the spatial space, the spatial translation invariance implies that, up to a normalization factor, $C_H(\vec{0})$ is the relative distribution density of $c$ at $\vec{r}$ relative to an antiquark $\bar{c}'$ at the origin. Consequently, for a bound state,of a $c\bar{c}'$ component, $C_H(\vec{r})$ can be viewed as the squared wave function that describes the spatial shape of $|H\rangle$. In this Letter, we focus on the spatial structures of the $1S$ ($\eta_c$ and $J/\psi$) and $1P$ ($h_c$ and $\chi_{c0,1,2}$) charmonia, as well as charmoniumlike states $\eta_{c1}$ and $h_{c0}$ of quantum numbers $J^{PC}=1^{-+}$ and $0^{+-}$, respectively. 

\paragraph{Numerical results---}
In the lattice QCD framework, the density-density correlation $C_H(\vec{r})$ is encoded in the four-point function
\begin{equation}
    C^{(4)}_H(t_1,t;\vec{r})=\int d^3 x\langle \mathcal{O}_H(t_1)\rho(\vec{x},t)\rho(\vec{x}+\vec{r},t)\mathcal{O}_H^\dagger(0)\rangle 
\end{equation}
where $\mathcal{O}_H$ is the zero-momentum projected interpolation operator with the same quantum numbers $J^{PC}$ as those of $|H\rangle$. Taking a spin-0 $|H\rangle$ for instance, one has
\begin{equation}\label{eq:ratio}
    C_H(\vec{r})\approx \frac{C^{(4)}(t_1,t;\vec{r})}{2m_H C^{(2)}_H(t_1)}\quad (t_1, t_1-t, t\gg 0),
\end{equation}

with $C^{(2)}_H(t)=\langle \mathcal{O}_H(t)\mathcal{O}_H^\dagger(0)\rangle$. The contamination from excited states is expected to be suppressed by taking this ratio. The spatial indices of $\mathcal{O}_H$ and the polarizations should be considered for a spin non-zero $|H\rangle$ state. 

The calculation of $C_H^{(4)}(t)$ and $C_H^{(2)}(t)$ is performed on a gauge ensemble with $N_f=2$ degenerate $u,d$ quarks. This gauge ensemble is generated on an $N_s^3\times N_t=16^3\times 128$ isotropic lattice with the aspect ratio $a_s/a_t\approx 5$ with $a_s$ and $a_t$ being the lattice spacing in the spatial and temporal directions, respectively. The spatial lattice spacing $a_s$ is determined to be 0.136(2) fm through the gradient flow method, and the light quark mass parameter is tuned to give a pion mass $m_\pi\approx 420$~MeV~\cite{Li:2024pfg,Liang:2024lon}. The mass parameter of the valence charm quark is tuned to give the physical spin-averaged mass of the $1S$ charmonium, namely, $\bar{m}(1S)=(3m_{J/\psi}+m_{\eta_c})/4=3096$ MeV.
\begin{figure}[t]
    \centering
    \includegraphics[width=0.46\linewidth]{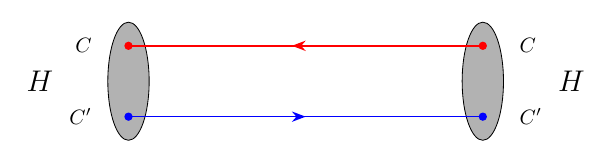}
    \label{fig:cr1}
    \includegraphics[width=0.46\linewidth]{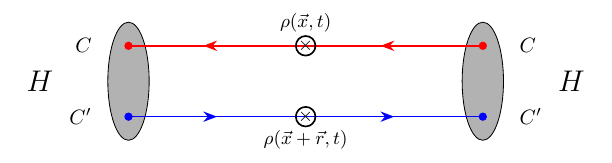}
    \label{fig:cr2}
    \hfill
    \caption{\justifying
    Schematic quark diagrams for correlation functions $C_H^{(2)}$ (left) and $C_H^{(4)}$ (right).
    }
    \label{fig:C4}
\end{figure}

If we assume $|H\rangle$ is a $c\bar{c}'$ bound state, then we consider only the quark diagrams in Fig.~\ref{fig:C4} for $C^{(2)}_H$ (left) and $C_H^{(4)}$ (right), respectively, which are dealt with using the distillation method~\cite{HadronSpectrum:2009krc}. The diagram on the left involves the perambulator of charm quark (the arrowed lines), namely, the all-to-all propagator that is distilled into the Heaviside subspace of the gauge invariant Laplacian operator on the lattice. Diagram (a) involves the generalized perambulator of charm quark, which is similar to the perambulator but with a local operator insertion $\rho(\vec{x},t)$ at $(\vec{x},t)$~\cite{Shultz:2015pfa}. The (generalized) perambulators are independent of states and can be calculated once for all. The operator $\mathcal{O}_H$ takes the quark bilinear form $\mathcal{O}_H^{(i)}=\bar{c}\Gamma_H^{(i)} c'$, where the quark fields $c$ and $c'$ are understood as the Laplacian-Heaviside smeared field, and $\Gamma_H^{(i)}$ refers to $\{\gamma_5, \gamma_i, 1,\epsilon_{ijk}\sigma_{jk}, \gamma_5\gamma_i, |\epsilon_{ijk}|\gamma_jD_k, \gamma_4, \epsilon_{ijk} \gamma_j B_k^{ab}\}$ for $\{\eta_c,J/\psi,\chi_{c0}, h_c,\chi_{c1},\chi_{c2},\eta_{c1}, h_{c0}\}$, respectively, with $D_k$ the lattice covariant derivative and $B_k^{ab}$ the chromomagnetic field strength. In practice, we choose $t_1=24 a_t$ and $t=t_1/2$ in calculating $C_H(\vec{r})$ through Eq.~(\ref{eq:ratio}), which are tested to be large enough for the suppression of the contamination from excited states. On the other hand, we use $\mathcal{O}_H^{(i=3)}$ (the $z$-component) to generate the state $|H\rangle$ for spin non-zero particles. This implies that the spin state is $|JJ_z\rangle=|10\rangle$ for $(J/\psi, h_c, \chi_{c1})$ and $\propto (|22\rangle-|2-2\rangle)$ for $\chi_{c2}$ (in the continuum limit).   
\begin{figure}[t]
    \includegraphics[width=0.32\linewidth]{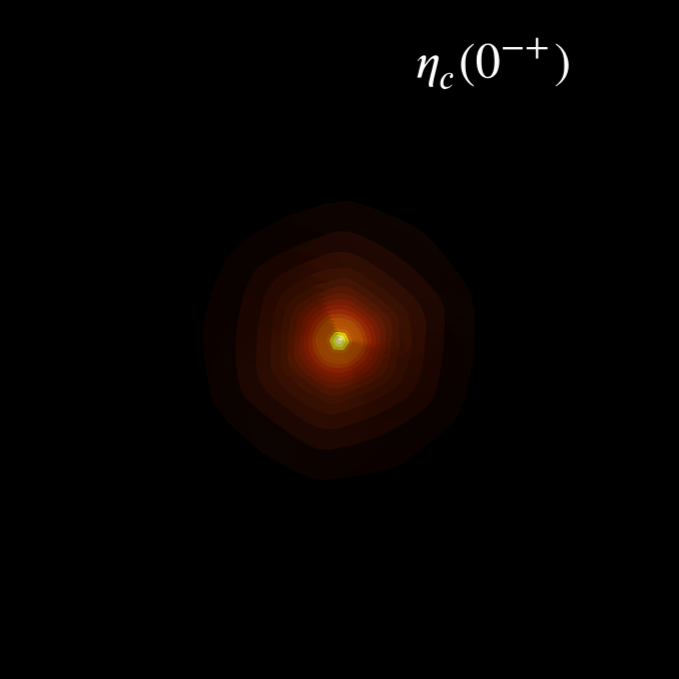}
    \includegraphics[width=0.32\linewidth]{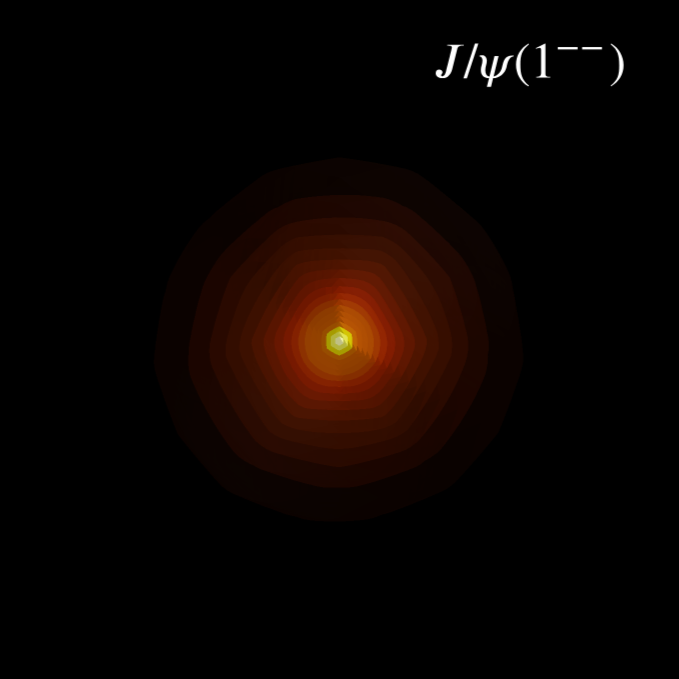}
    \includegraphics[width=0.32\linewidth]{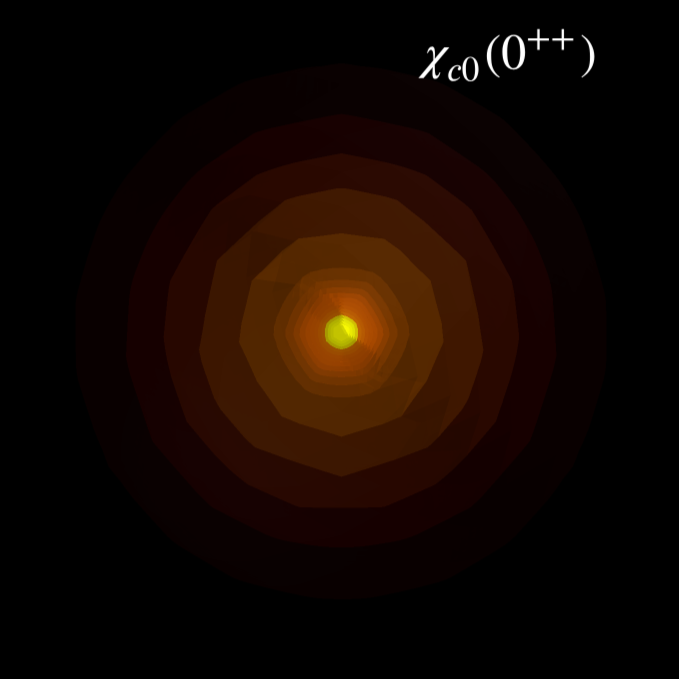}
    \includegraphics[width=0.32\linewidth]{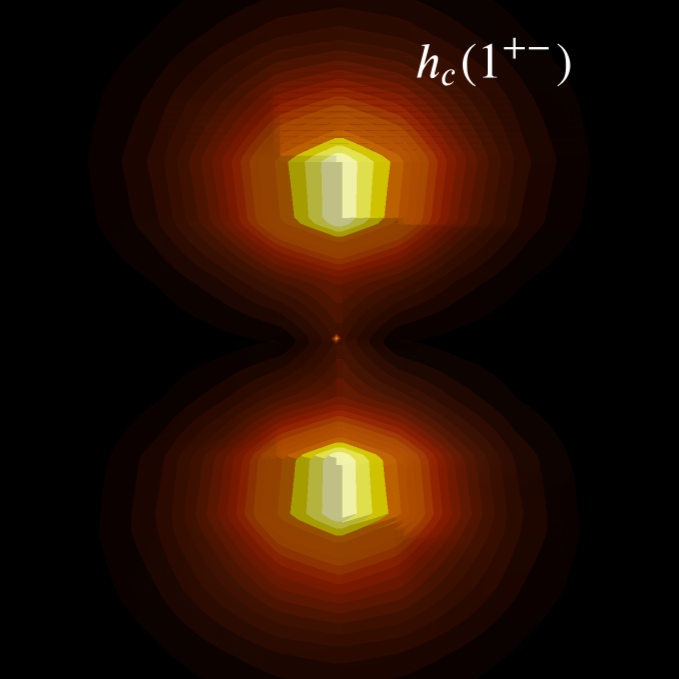}
    \includegraphics[width=0.32\linewidth]{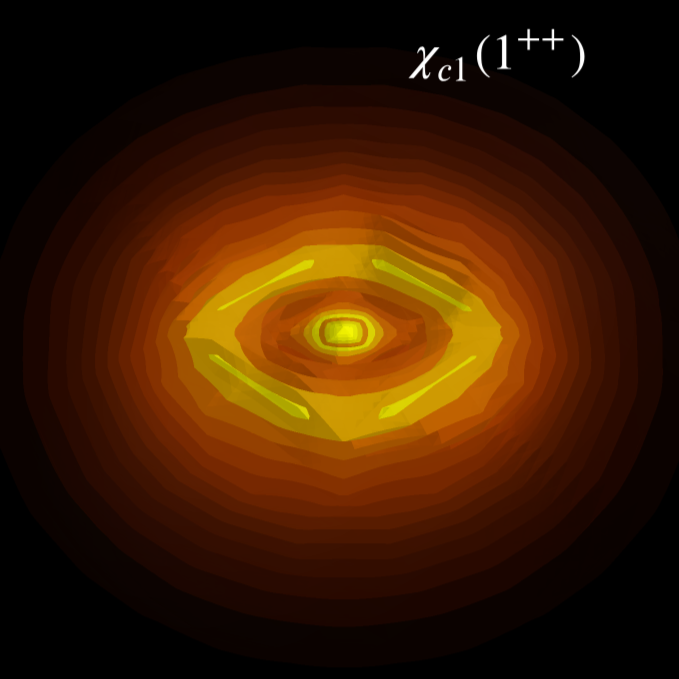}
    \includegraphics[width=0.32\linewidth]{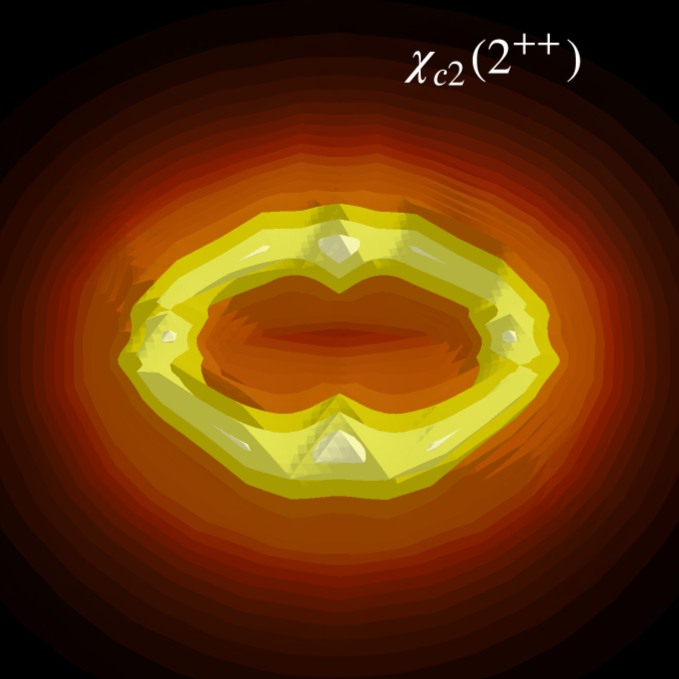}
    \vskip 0.2cm
    \includegraphics[width=0.45\linewidth]{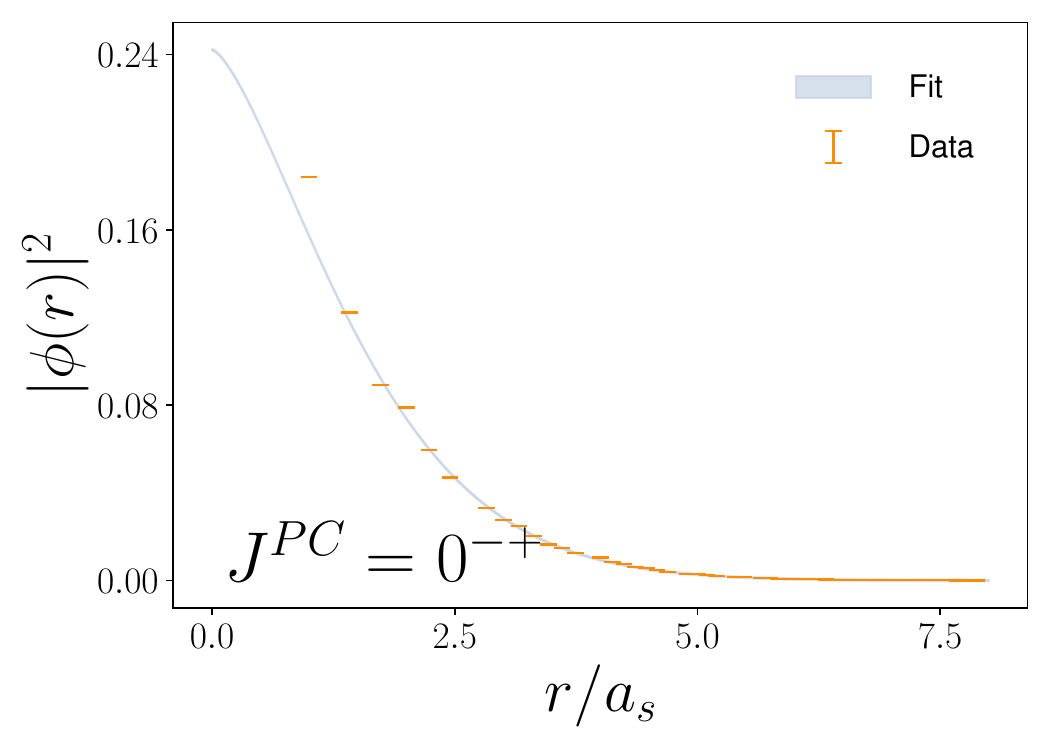}
    \includegraphics[width=0.45\linewidth]{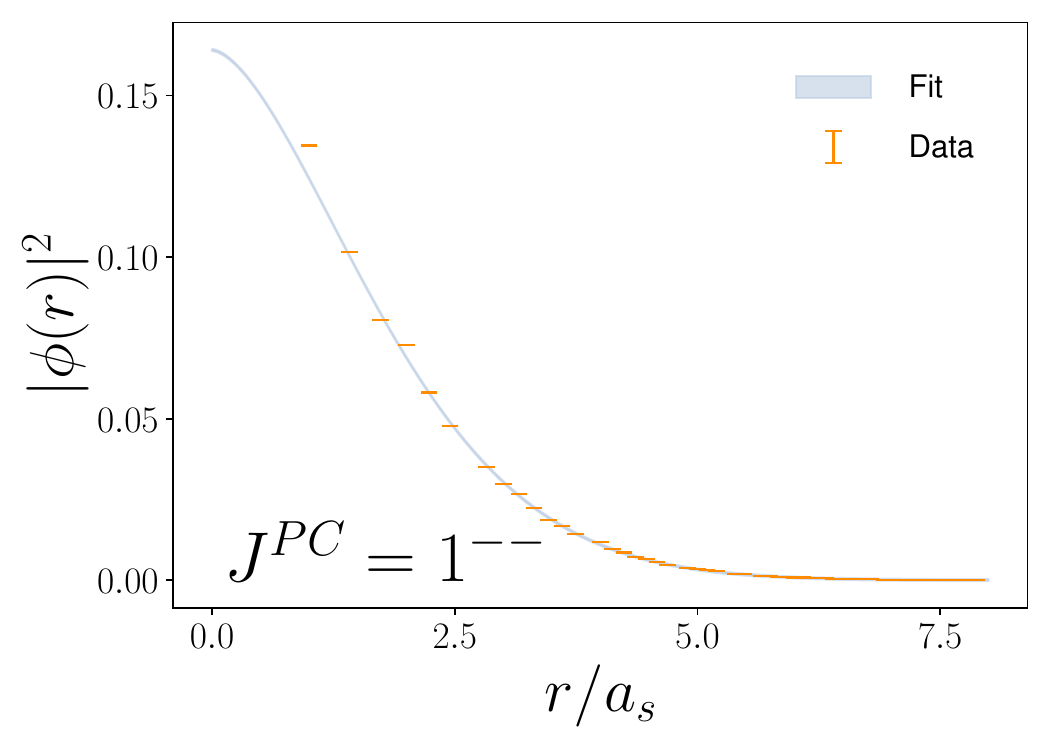}\\
    \includegraphics[width=0.45\linewidth]{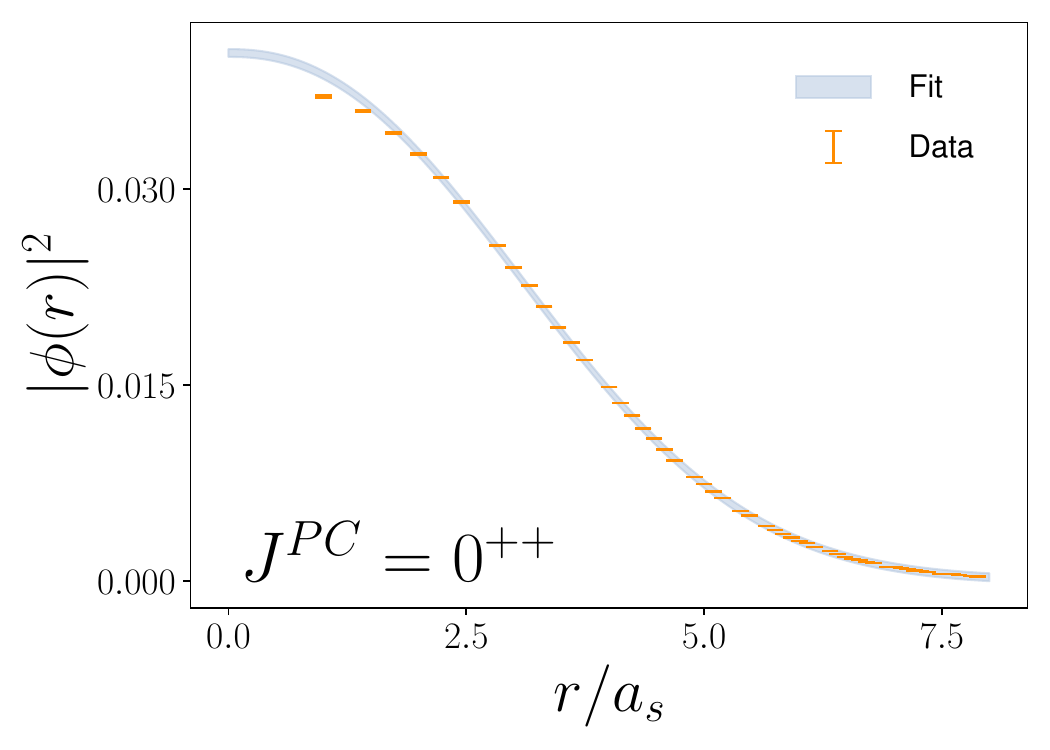}
    \includegraphics[width=0.45\linewidth]{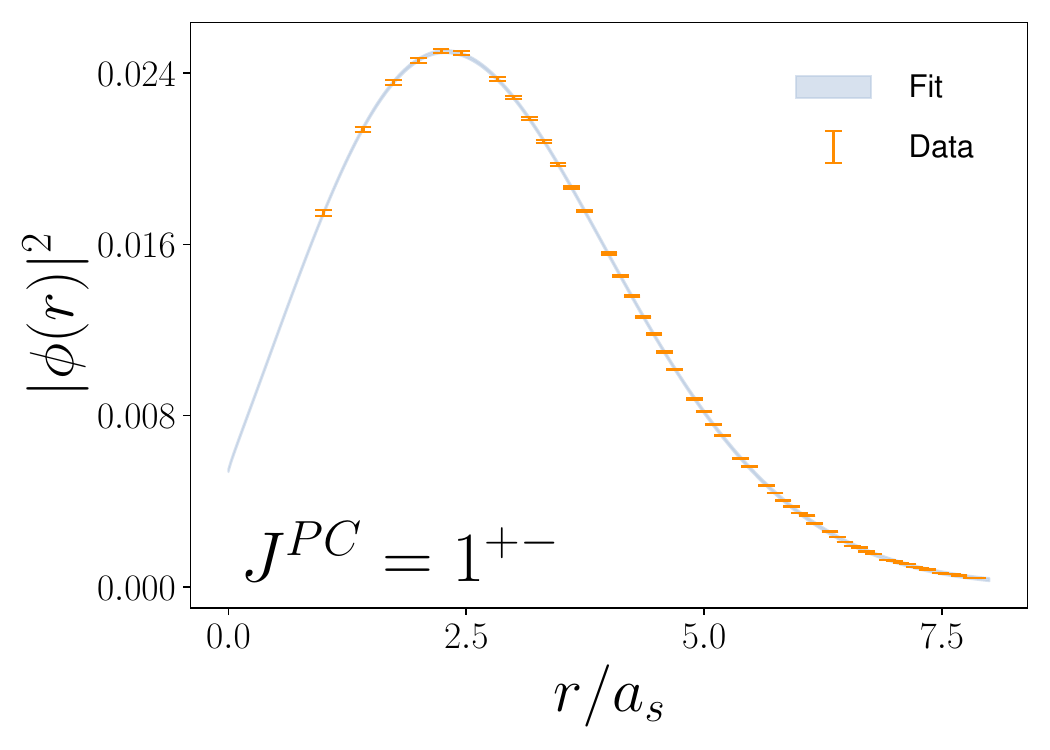}\\
    \includegraphics[width=0.45\linewidth]{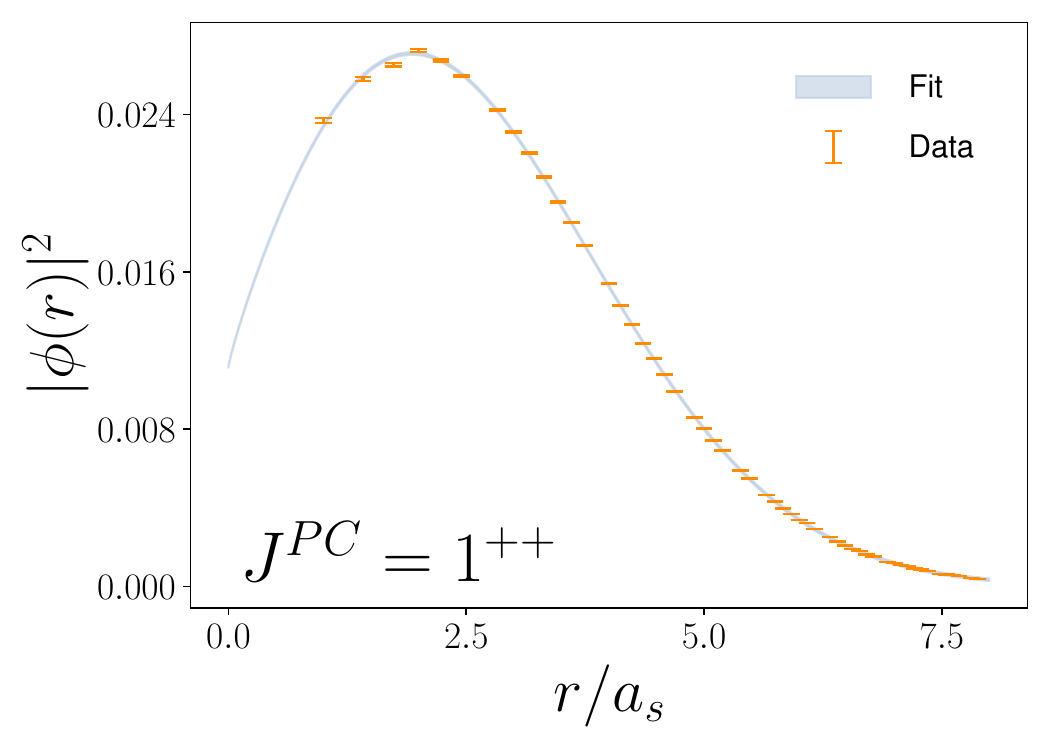}
    \includegraphics[width=0.45\linewidth]{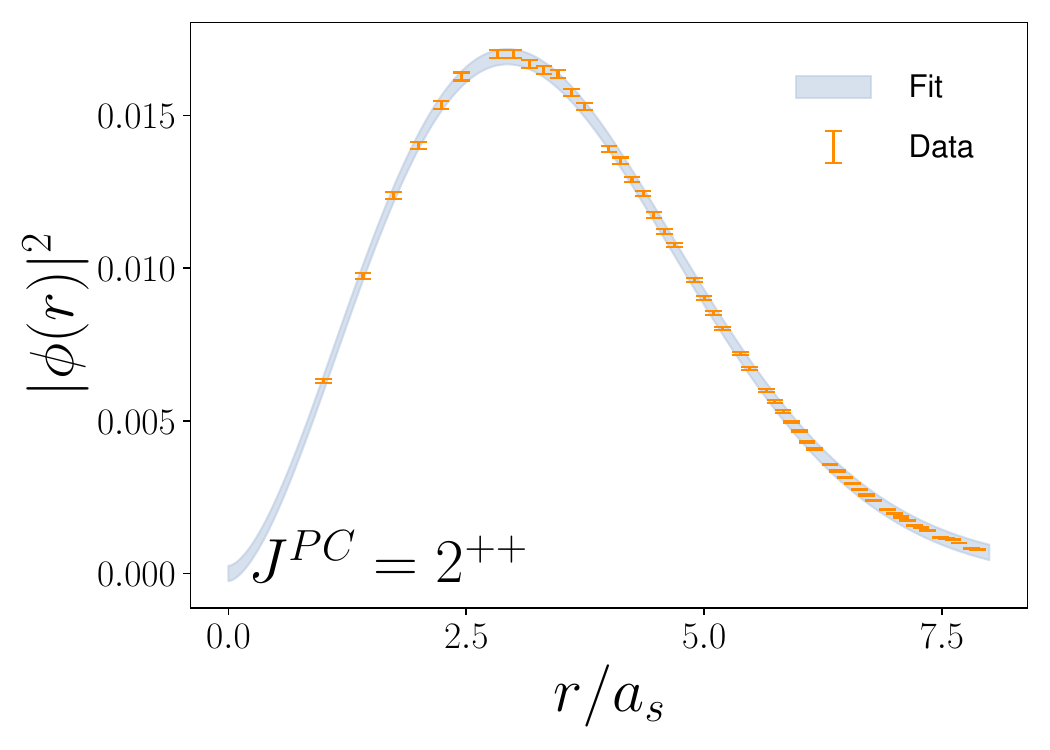}
    \caption{\justifying
    Spatial distributions $C_H(\vec{r})$ for $1S$ and $1P$ charmonia. {\bf The upper six panels:} The $z$-axis is vertical and passes through the center of the distribution. 
    All panels share a common logarithmic colour scale and cover the same spatial volume. 
    {\bf The lower six panels:}  Data points show the radial distributions $|\phi_H(r)|^2$ and the shaded curves show the fits through function forms $A e^{-(r/r_0)^\alpha} + B r e^{-(r/r_0)^\beta}$ excluding the values at $r=0$.
    \label{fig:3-d}
    }
\end{figure}
\paragraph{Portraits of $1S$ and $1P$ charmonia---}
The spatial distributions $C_H(\vec{r})$ for $1S$ and $1P$ states are shown in Fig.~\ref{fig:3-d}, where the $z$-axis is vertical and passes through the center of the distribution. The angular distributions of $1S$ states ($\eta_c$ and $J/\psi$) are spherical as expected from NRQMs which describes them to be the $L=0$ states. The charmonium states $h_c(1^{+-})$, $\chi_{c0}(0^{++})$, $\chi_{c1}(1^{++})$ and $\chi_{c2}(2^{++})$ are usually assigned by NRQMs to be $1P$ states, namely, $n^{2S+1}L_J=1^{1}P_1$ and $1^3 P_{0,1,2}$ states, respectively. The spherical shape of $C_H(\vec{r})$ for $\chi_{c0}(0^{++})$ is natural and is resulted from the spin-orbital coupling. In order to understand the angular distribution of other $1P$ states, we factorize $C_H(\vec{r})$ into the product of the radial part and the angular part, namely,
\begin{equation}
    C_H(\vec{r})=|\phi_H(r)|^2 |\xi_H(\theta,\phi)|^2.
\end{equation}
NRQMs expect the $|JJ_z\rangle=|10\rangle$ states of $h_c(L=1,S=0)$ and $\chi_{c1}(L=1,S=1)$ have angular distributions $|\xi_H(\theta,\phi)|^2\propto \cos^2\theta$ (spindle shape) and $\sin^2\theta$ (donut shape), respectively, and that for the $(|22\rangle+|2-2\rangle)$ state of $\chi_{c2}$ is proportional to $ \sin^2\theta$. These are more or less the cases for the measured results of $|\xi_H(\theta,\phi)|^2$ shown in Fig.~\ref{fig:3-d}. We have tested rotational symmetry for $z\to x\to y$ transformation. However, there is also a clear $\phi$-dependence of $|\xi_H(\theta,\phi)|^2$ for $\chi_{c1,2}$, and $C_H(r)$ is non-zero for $h_c$ and $\chi_{c0,1}$. These phenomena are unexpected by NRQMs where $1P$ states are pure $L=1$ states with $\phi_H(r=0)=0$, and are attributed to the relativistic effect (see below).

    \begin{table*}[htbp]
        \renewcommand\arraystretch{1.5}
        \caption{\justifying
            From top to bottom, the RMS radii $\sqrt{\langle r^2\rangle}$, the lattice-extracted wave-function values at the origin $P(0)$, and the fitted origin values $|\phi(0)|^2$ for various charmonium-like states $H(J^{PC})$. The difference between $P(0)$ and $|\phi_H(0)|^2$ signals the weak divergence at $r=0$. $P(0)\approx |\phi_H(0)|^2\approx 0$ is exactly the same as expectation for $\chi_{c2}$. }
        \label{tab:size}
        \begin{ruledtabular}
            \begin{tabular}{ccccccccc}
                $H(J^{PC})$
                                                 & $\eta_c(0^{-+})$ & $J/\psi(1^{--})$ & $h_c(1^{+-})$ & $\chi_{c0}(0^{++})$ & $\chi_{c1}(1^{++})$ & $\chi_{c2}(2^{++})$ & $\eta_{c1}(1^{-+})$ & $h_{c0}(0^{+-})$ \\
                                                                 $P(0)$                           & 0.3973(4)        & 0.2124(2)        & 0.0277(2)     & 0.0891(2)           & 0.0464(2)           & 0.00176(7)          & 0.0459(39)  & 0.0169(36)  \\
                $|\phi_H(0)|^2$                  & 0.2421(2)        & 0.1641(1)        & 0.0202(54)    & 0.04042(5)          & 0.0371(12)          & 0.0000(2)           & 0.0451(21)  & 0.0161(83)\\
                $\sqrt{\langle r^2\rangle}$ (fm) &
                0.423(1)                    & 0.441(2)         & 0.609(4)         & 0.585(4)      & 0.604(4)            & 0.669(6)            & 0.582(71)           & 0.588(84)                
            \end{tabular}
        \end{ruledtabular}
    \end{table*}
After integrating over (sum over on the lattice) the angular directions, we obtain the radial distribution densities $|\phi_H(r)|^2$ for different states $|H\rangle$, which are shown in the lower six panels in Fig.~\ref{fig:3-d} with the normalization factor $4\pi \int dr r^2|\phi_H(r)|^2$. For $1S$ states ($\eta_c$ and $J/\psi$), 
the radial behaviors are in qualitative agreement with the expectation of the non-relativistic $1S$ bound states. For $1P$ states except for $\chi_{c2}$, 
$|\phi_H(r)|^2$'s do not meet the NRQM prediction that $|\phi_H(r)|^2$ take the asymptotic behavior $|\phi_H(r)|^2\sim r^2$ for $r\to 0$. 

These phenomena can be understood in the Dirac theory for a bound state of two spin-1/2 particles. Considering a charm quark $c$ in a central potential generated by the charm anti-quark ($c'$). We focus on the $r\sim 0$ region where the one-gluon-exchange potential $-\alpha_s/r$ dominates the interaction and the confining potential can be neglected. Then the Dirac equation $(\vec{\alpha}\cdot \hat{p}+\beta m_c)\Psi=(E+\alpha_s/r)\Psi$ ($\beta=\gamma^0$ and $\vec{\alpha}=\beta\vec{\gamma}$) of the charm quark can be solved similarly to that of the electron in the hydrogen atom~\cite{Greiner:1997xwk}. For the charm quark in this system, its total angular momentum $\hat{j}=\hat{L}+\mathbf{\Sigma}_c/2, \hat{j}_z$, the quantity $\hat{K}=\gamma_0 (1+\mathbf{\Sigma}_c\cdot \hat{L}$ )(Here $\mathbf{\Sigma}_c=\left(\begin{array}{cc}\vec{\sigma}&0\\0&\vec{\sigma}\end{array}\right)$ is the spin matrix in the spinor space) and $\hat{\alpha}=\vec{\alpha}\cdot \vec{r}/r$ share the common eigenfunction with the eigenvalues $j,m,\kappa=\pm(j+1/2),\alpha_r=\pm 1$, respectively, which is also the solution to the Dirac's equation and takes the form
    \begin{equation}
        \Psi_{n,\kappa,m}(\vec{r})=\begin{pmatrix}F_{n,\kappa}(r)\chi_{j,m}^{(l)}(\theta,\phi)\\iG_{n,\kappa}(r)\chi_{j,m}^{(l')}(\theta,\phi)\end{pmatrix}.
    \end{equation}
where $n$ is the principle quantum number ($n=1$ for $1S$ states and $n=2$ for $1P$ states) ( $\chi_{j,m}^{(l)}$ is a Pauli spinor with $l'=l+1,\kappa=-l'$ for $j=l+1/2$ and $l'=l-1, \kappa=l$ for $j=l-1/2$, and the radial wave functions are normalized as $\int r^2dr (G^2+F^2)=1$ (see Supplemental Material for details). The wave function $\Psi_{n,\kappa,m}(\vec{r})$ has a definite parity $(-)^l$.  We would like to emphasize that the upper and lower component of $\Psi_{n,\kappa,m}(\vec{r})$ are eigenstates of $\hat{L}$ with eigenvalues $l$ and $l'=l\pm 1$, respectively. Note that $\hat{\alpha}_r$ actually reflects the (chromo-)electric dipole interaction owing to the relativistic effect, and mixes the upper and the lower components of $\Psi_{n,\kappa,m}(\vec{r})$ and therefore the orbital angular momentum states of $l$ and $l\pm 1$. 
    
According to the explicit solution in Ref.~\cite{Greiner:1997xwk}, the radial wave function has the asymptotic behavior near $r=0$, 
    \begin{equation}
        \left(F_{n,\kappa}(r),G_{n,\kappa}(r)\right)\sim \left(\sqrt{m_c+E_{n,\kappa}},\sqrt{m_c-E_{n,\kappa}}\right) r^{s-1}
    \end{equation}
with $s=\sqrt{\kappa^2-\alpha_s^2}$. For lowest lying charmonium states and neglecting the confining potential, one has $m_c-E_{n,\kappa}\sim \mathcal{O}(m_c\alpha_s^2)$, such that the lower component is important and has physical consequences, since $G_{n,\kappa}/F_{n,\kappa}\sim \mathcal{O}(\alpha_s)$ for the typical value $\alpha_s\sim 0.3$. This is in striking contrast to the case of hydrogen atom where $\alpha\approx 1/137$.    
    
The bound state condition requires $F(r)$ and $G(r)$ behave as $(F,G)(r)\sim r^{s-1}$ when $r\to 0$, and the mixing of $l$ nad $l\pm 1$ result in different angular distributions from the expectation of NRQMs. The details can be found in Ref.~\cite{Greiner:1997xwk} and the Supplemental Materials. Here we only quote the major features as follows:
\begin{itemize}
    \item $\eta_c$ and $J/\psi$: Their quantum numbers are $(n,j,\kappa)=(1,1/2,-1)$, and $F_{1,-1}(r)$ and $G_{1,-1}$ has the same asymptotic behavior $\sim r^{s-1}\approx r^{-\alpha_s^2}/2$ when $r\to 0$. The orbital part of the lower component has $l'=1$, but when considering the spin of the antiquark $\bar{c}'$, the angular distribution is actually spherical. 
    \item $\chi_{c0}$: The quantum number is $(n,j,\kappa)=(2,1/2,1)$, and the asymptotic $r$ behaviors of $F_{2,1}$ and $G_{2,1}$ are $\sim r^{s-1}(\mathcal{O}(\alpha_s^2)+r)$ and $r^{s-1}$, respectively, when $r\to 0$. So the $r$ behavoir of $C_H(\vec{r})$ is dominated by the lowere component. The angular distribution of the upper component is similar to the NRQM expectation, while that of the lower one is also spherical since $l'=0$. 
    \item $h_c$: It is an admixture of $(n,j,\kappa)=(2,1/2,1)$ and $(2,3/2,-2)$ states. Due to the total spin $S=0$, only the upper components contribute to the spatial distribution. So it is a purely $l=1$ state with the nonzero value of $C_H(\vec{r})$ at at $r=0$ from the tiny value of $F_{2,1}(r=0)$. 
    \item $\chi_{c1}$: It is also an admixture of $(n,j,\kappa)=(2,1/2,1)$ and $(2,3/2,-2)$ states. The lower components of $\Psi_{n,\kappa,m} (\vec{r})$ contribute to the spatial distribution and have $l'=0$ and $l'=2$ orbital parts. So $G_{2,1}(r)\sim r^{s-1}$ results in a large nozero value near $r=0$. The $l=2$ component makes the angular distribution shows a weak quadrupole shape. 
    \item $\chi_{c2}$: It is a pure $(n,j,\kappa)=(2,3/2,-2)$ state. The lower component has a $l'=2$ orbital part which results in a clear quadrupole angular distribution. The samll $r$ behavior is due to $(F_{2,-2},G_{2,-2})(r)\sim r^{s-1}$ which is exactly zero at $r=0$.  
\end{itemize}
For all the $1S$ and $1P$ states except for $\chi_{c2}$, the radial distribution behaves $\sim r^{-\alpha_s^2}$ near $r=0$ due to the $(F_{n,\pm 1},G_{n,\pm 1})$ contribution, which diverges for a sizable $\alpha_s^2\sim 0.1$. We do observe this divergence in $C_H(\vec{r})$ for these states, as shown in Fig.~\ref{fig:3-d}. We try to describe the radial distribution $|\phi_H(r)|^2$ using the function form like $|\phi_H(r)|^2=A e^{-(r/r_0)^\alpha} + B r e^{-(r/r_0)^\beta}$ with the exclusion of value of $C_H(\vec{r}$ at $r=0$ and obtain fairly good fits, as shown in the lower group of Fig.~\ref{fig:3-d} by colored bands. The comparison of the fitted $|\phi_H(0)|^2$ and the measured values labelled by $P(0)$ (with the same normalization) is shown in Table~\ref{tab:size} and serves as a quantitative measure of the $r=0$ divergence to some extent. 

On the other hand, we can estimate the root-mean-squared (RMS) `radius' $r_\mathrm{RMS}\equiv \sqrt{\langle r^2\rangle}$ of the charmonium(like) state $H$ with $\langle r^2\rangle$ being defined as
    \begin{equation}\label{eq:rms}
        \langle r^2\rangle=\frac{\int d^3\vec{r}~ r^2 C_H(\vec{r})}{\int d^3\vec{r}~  C_H(\vec{r})}.
    \end{equation}
The value of $r_\mathrm{RMS}$ measures the size of a conventional charmonium state or the means separation between charm quark and antiquark in a charmoniumlike state with additional degree(s) of freedom. The values of $r_\mathrm{RMS}$ for different states $H$ are collected in Table~\ref{tab:size}. The sizes of $\eta_c$ and $J/\psi$ are 0.423(1) fm and 0.441(2) fm, respectively. The sizes of $h_c, \chi_{c0}, \chi_{c1}$ are around 0.6 fm and are a little smaller than that of $\chi_{c2}$. It is interesting to see that the non-relativistic quark model predictions~\cite{Wong:2001td,Akbar:2011jd} are compatible with our predictions.

The precise derivation of the radial functions $F(r)$ and $G(r)$ can be done numerically or theoretically with a given potential $V(r)$, but this is out of the scope of this work. We would like to emphasize that, the spatial distributions of charm quark and antiquark in $1P$ charmonia are very different from the non-relativistic quark model (even plus relativistic corrections), and the relativistic effects due to the quark spin are important for the dynamics of the charmonium formation.

    \paragraph{$1^{-+}$ and $0^{+-}$ charmonium like states---}The quantum numbers $J^{PC}=1^{-+}$ and $0^{+-}$ are exotic quantum numbers that are prohibited for quark-antiquark mesons. So in the quark model picture, a charmoniumlike state of these exotic quantum numbers should have additional light degrees of freedom. If they are gluonic, then the state is usually named as a $c\bar{c}g$ hybrid meson. 
    

    \begin{figure}[t]
        \includegraphics[width=0.3\linewidth]{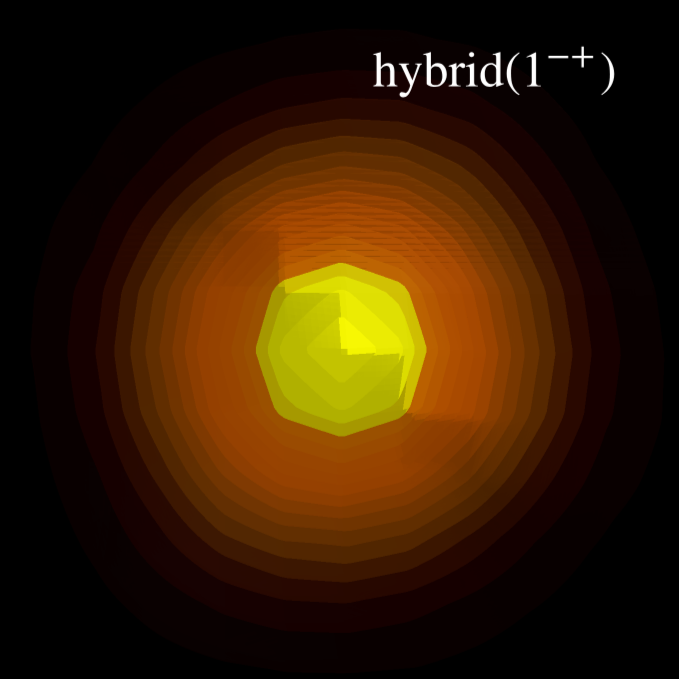}
        \includegraphics[width=0.5\linewidth]{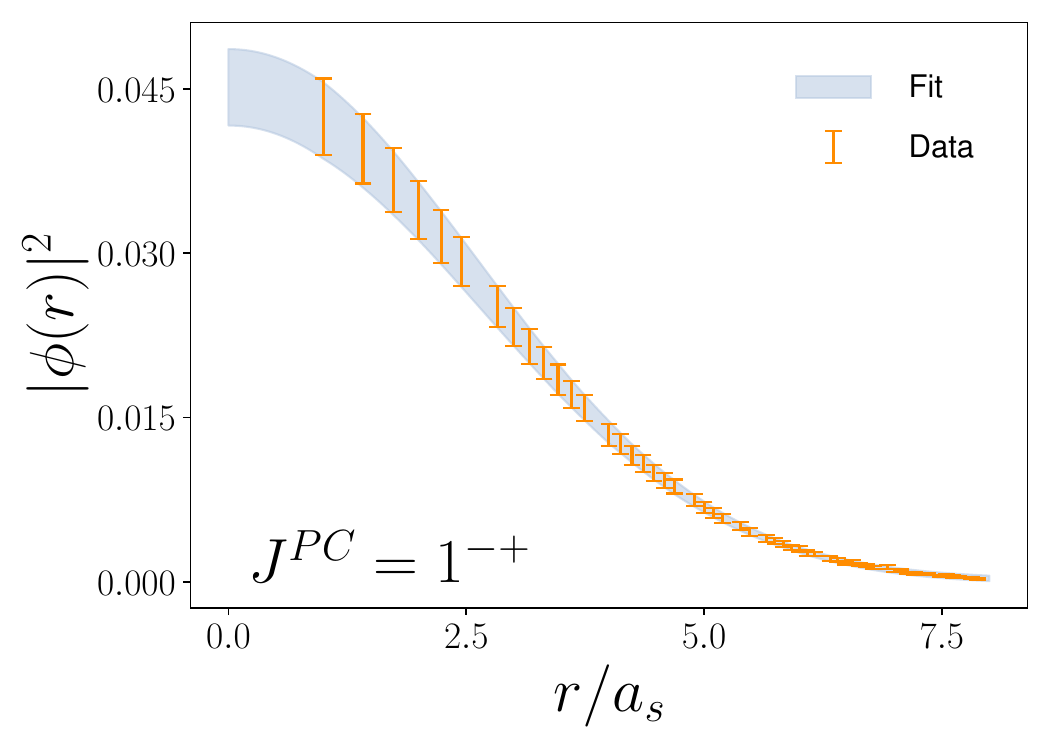}\\
        \includegraphics[width=0.3\linewidth]{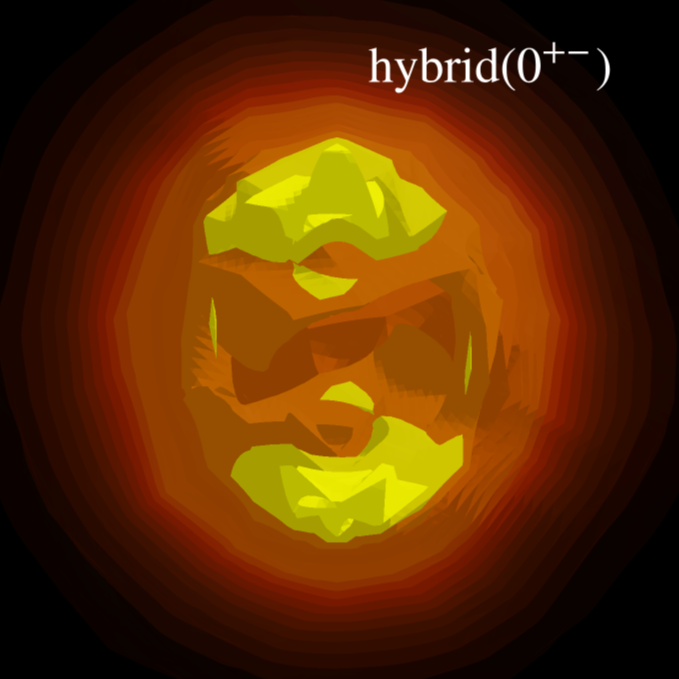}
        \includegraphics[width=0.5\linewidth]{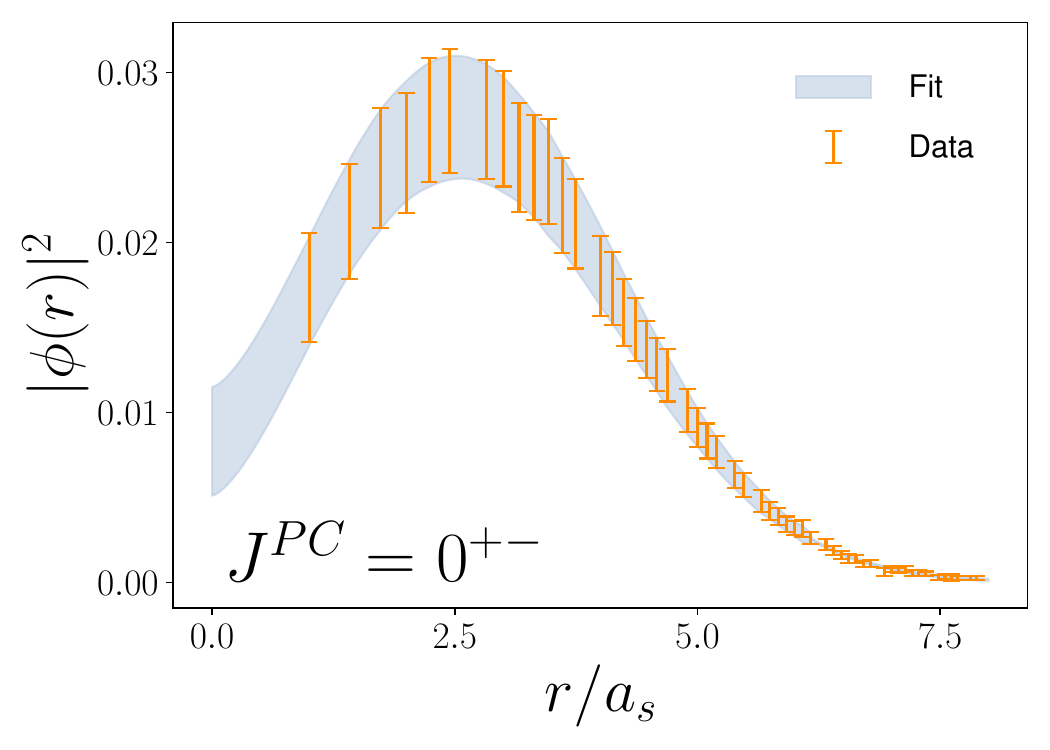}
        \caption{\justifying
            Three‑dimensional probability‑density renderings for the
            exotic charmonium‑like states, shown top‑to‑bottom as $1^{-+}$ and $0^{+-}$. Visualization conventions—spin axis ($z$–direction), colour
            scale, normalization, and spatial volume—are identical to Fig.~\ref{fig:3-d}.\label{fig:3-d-exotic}}
    \end{figure}
    
Figure~\ref{fig:3-d-exotic} illustrates the spatial distributions of $C_H(\vec{r})$ and their radial behaviors for $\eta_{c1}(1^{-+})$ and $h_{c0}(0^{+-})$. For a state of more than two components, there should be additional dynamical variables other than the spatial separation $\vec{r}$ between charm quark ($c$) and antiquark ($c'$). So $C_H(\vec{r})$ here can be viewed as the distribution density of $c$ relative to $\bar{c}'$ after integrating over other variables. We can also decompose $C_H(\vec{r})$ into the product of the radial and orbital parts, namely, $C_{H}(\vec{r})=|\phi_H(r)|^2|\xi_H(\theta,\phi)|^2$, to interpret their spatial distributions: $\eta_{c1}(1^{-+})$ has a spherical angular distribution $|\xi_H(\theta,\phi)|^2$ and a radial distribution $|\phi_H(r)|^2$ similar to those of $1S$ states or the $0^{++}$ charmonium $\chi_{c0}$. Given a $c\bar{c}g$ hybrid, the usual convention that the charge conjugate parity ($C$-parity) $C'(g)=-$ of a gluon requires the $C$-parity of the $c\bar{c}$ to be negative. Thus the angular distribution $|\xi_H(\theta,\phi)|$ implies the $c\bar{c}$ in this state has quantum numbers $J^{PC}=1^{--}$ and the $1^{+-}$ gluonic component and the $c\bar{c}$ component are in a relative $S$-wave~\cite{Dudek:2011bn,Dudek:2012ag,Ma:2019hsm}. This argument also applies to $h_{c0}(0^{+-})$ state which may has a $1^{+-}$ gluonic component and a $1^{++}$ $c\bar{c}$ component~\cite{Dudek:2011bn}. Its $|\phi_H(r)|^2$ does resemble that of $\chi_{c1}$, and Averaging over the polarization of the $1^{++}$ $c\bar{c}$ gives a spherical distribution. The values of $r_\mathrm{RMS}$ of $\eta_{c1}$ and $h_{c0}$ are also derived using Eq.~(\ref{eq:rms}) to be 0.58(7) fm and 0.59(8) fm, respectively, which are comparable with those of $1P$ states in Table~\ref{tab:size}.  

As to $\eta_{c1}(1^{-+}$) hybrids, a previous lattice QCD study also find that the $c\bar{c}$ component and the $1^{+-}$ gluonic component are in $S$-wave, which is reflected by the Coulomb gauge Bethe-Salpeter wave function with respect to their separation~\cite{Ma:2019hsm}. Since both components are color octets, the Casimir scaling of the static potential between two color charges~\cite{Bali:2000un} expects the confining dynamics of the two components can be described by the potential     
    \begin{equation}
        V(r)=V_0 + \frac{9}{4}\left(-\frac{4\alpha_s}{3r}+\sigma r\right),\label{mass}
    \end{equation}
where $V_0$ is a constant, $\sigma$ is the string tension of $Q\bar{Q}$ pair, the factor $9/4$ is the ratio of the second order Casimir operator of color octet ($C_2^{(8)}=3$) to that of the triplet ($C_2^{(3)}=4/3$). Considering the gluonic component moving in the potential generated by the $c\bar{c}$ component and solving the non-relativistic Schr\"{o}dinger equation, one obtains the mass splitting between the $1S$ and $2S$ states
$\Delta m(2S-1S)=((9\sigma/4)^2/(2\mu))^{1/3}(d_2-d_1)$, where $\mu$ is the reduced mass of the system, and $(d_1, d_2)$ are eigenvalues corresponding to $m(1S,2S)$~\cite{Eichten:1978tg}. If the effective masses of the $c\bar{c}$ block and the gluelump are taken to be 3.0 GeV and 0.8 GeV, then using the values $\alpha_s \approx 0.3$, $\sigma\approx 0.25\,\mathrm{GeV}^2$, one has
$(d_1=1.556, d_2=3.572)$, and thereby $m(2S-1S)\sim 1.272$ GeV, which is in fairly good agreement with the value 1.1-1.3 GeV obtained from lattice QCD calculation~\cite{Ma:2019hsm}.    

It is stressed that the observations in this Letter and the previous lattice studies on $1^{-+}$ charmoniumlike hybrids are strikingly different from the picture in the Born-Oppenheimer approximation that the gluonic degrees of freedom are taken light and fast ones which move around the axis along the $c\bar{c}$ pair and result in a hybrid confining potential~\cite{Juge:1999ie,Braaten:2014qka}. In this picture, the $c\bar{c}$ of 
$\eta_{c1}(1^{-+}$ is in an effective $P$-wave and the first excitated state ($2P$ state) is higher than the ground state ($1P$) by around 350 MeV, which is much smaller than the lattice result~\cite{Ma:2019hsm} and the estimate in this Letter.

\paragraph{Summary---} The internal structures of charmoniumlike states are explored through the charge density-density correlations within these states, which are gauge invariant and model independent. The observed spatial distributions of $c\bar{c}$ in $1S$ and $1P$ charmonia exhibit important relativistic effects and can be understood neatly in the Dirac theory of quarks where quark spin is naturally incorporated. For the charmoniumlike hybrid $\eta_{c1}(1^{-+})$, the relative distribution of $c\bar{c}$ indicates that its $c\bar{c}$ component has quantum numbers $1^{-(-)}$, such that $\eta_{c1}$ can be view as a bound state of a color octet $c\bar{c}$ and a $1^{+-}$ gluonic (chromomagnetic) componet in $S$-wave. This spatial configuration of $\eta_{c1}$ does not support the flux tube picture in the Born-Oppenheimer approximation. The findings in this Letter provide {\it ab initio} information for understanding the formation of charmoniumlike states.

\begin{acknowledgments}
    \paragraph{Acknowledgments.---}
    This work is supported by the National Natural Science Foundation of China (NNSFC) under Grants No. 12293060, No. 12293065. WS is also supported by Chinese Academy of Sciences under Grant No.\,YSBR-101. The Chroma software system\,\cite{Edwards:2004sx}, QUDA library\,\cite{Clark:2009wm,Babich:2011np},
    and PyQUDA package\,\cite{jiang2024usequdalatticeqcd} are acknowledged. The computations were performed on the HPC clusters at the Institute of High Energy Physics (Beijing), China Spallation Neutron Source (Dongguan), and the ORISE computing environment.
\end{acknowledgments}

\bibliography{ref}
\clearpage

\onecolumngrid
\section{Supplementary Materials}
\setcounter{equation}{0}
\setcounter{figure}{0}
\setcounter{table}{0}
\renewcommand{\theequation}{S\thesection\arabic{equation}}
\renewcommand{\thefigure}{S\thesection\arabic{figure}}
\renewcommand{\thetable}{S\thesection\arabic{table}}

\subsection{S1. Relativistic angular‑momentum coupling: A complete derivation}\label{sec:correction}
When considering relativistic effects in the charmonium system, it becomes necessary to incorporate quark spin interactions by replacing the non-relativistic Schrödinger equation with the Dirac equation. This transition is expected to induce significant corrections in the radial wavefunction—not merely modifying the angular components—thereby reflecting the more complex structure of the relativistic dynamics. One can refer the textbook of Ref.~~\cite{Greiner:1997xwk} for the details of the solution to the Dirac theory of hydrogen.

Consider a charm quark in the gluonic field generated by the charm antiquark and ignore the their annihilation effect, and assume the field is described by the usually used Cornell potential
\begin{equation}
    V(r)=V_0-\frac{\alpha_s}{r}+\sigma r.
\end{equation}
We are interested in the radial wave function of the charm quark near the origin, say, $r\sim 0$. So the effect of the linear part can be ignored temporarily and the problem is exactly the same as an electron in the Coulomb potential. The total angular momentum $\hat{j}=\hat{L}+\mathbf{\Sigma_c}/2$, its projection in the $z$-direction $\hat{j}_z$, and the quantity
\begin{equation}
    \hat{K}=\gamma_0 \left(1+\Sigma_c\cdot \hat{L}\right), ~~\Sigma_c=\mathbf{\Sigma}_c=\left(\begin{array}{cc}\vec{\sigma}&0\\0&\vec{\sigma}\end{array}\right)
\end{equation}
establish a complete quantity set and share the same eigenfunction (here we using the Dirac representation)
\begin{equation}
    \Psi_{n,\kappa,m}(\mathbf{r})=\begin{pmatrix}F_{n,\kappa}(r)\varphi_{\kappa,m}(\theta,\phi)\\iG_{n,\kappa}(r)\varphi_{-\kappa,m}(\theta,\phi)\end{pmatrix}.
\end{equation}
labelled by the eigenvalues $n,j,\kappa,m$ that are defined as
\begin{eqnarray}
    (\vec{\sigma}\cdot \hat{L}+1)\varphi_{\kappa,m}&=&-\kappa \varphi_{k,m}\nonumber\\
    (\vec{\sigma}\cdot \hat{L}+1)\varphi_{-\kappa,m}&=&\kappa \varphi_{-\kappa,m}\nonumber\\
    \hat{j}_z \varphi_{\kappa,m}&=&m \varphi_{\kappa,m},
\end{eqnarray}
and $\kappa$ takes the values $\pm 1, \pm 2, \pm 3,\ldots$, and is related to the eigenvalue of $\hat{j}$ by $\kappa=\pm (j+1/2)$. The quantum number $n$ is the so-called principle quantum number and will be explained after a while. It can easily verified that $\Psi_{n,\kappa,m}(\vec{r})$ is also the eigen function of the Hamiltonian $\hat{H}$ of the Dirac theory,
\begin{equation}
    \left(\vec{\alpha}\cdot \hat{p}+\beta m_c\right)\Psi_{n,\kappa,m}=\left(E_{n,\kappa}+\frac{\alpha_s}{r}\right)\Psi_{n,\kappa,m}(\vec{r}),
\end{equation}
where $m_c$ is the effective mass of the charm quark, the potential constant $V_0$ is absorbed in $E_{n,\kappa}$, and $\beta=\gamma_0$ and $\vec{\alpha}=\beta \vec{\gamma}$ have been defined. Since the Hamiltonian is invariant under the parity transformation $\mathcal{P}$, the eigen function
$\Psi_{n,\kappa,m}(\mathbf{r})$ should have a definite eigenvalue, namely
\begin{equation}
    \mathcal{P}\Psi_{n,\kappa,m}(\vec{r})=\beta \Psi_{n,\kappa,m}(-\vec{r})=(-)^l \Psi_{n,\kappa,m}(\vec{r})\equiv (-)^l \phi_{ljm},
\end{equation}
where $l$ is the orbital momentum of the charm quark. Thus the parity of the charmonium state is $P'=(-)^{l+1}$. The solution that meets this requirement is expressed as
\begin{equation}
    \begin{aligned}&\chi_{j,m}^{(l=j-1/2)}=\left(2l+1\right)^{-1/2}\begin{pmatrix}\left(l+m+\frac{1}{2}\right)^{1/2}Y_{l,m-1/2}\\(l-m+\frac{1}{2})^{1/2}Y_{l,m+1/2}\end{pmatrix}\\&\chi_{j,m}^{(l=j+1/2)}=(2l+1)^{-1/2}\begin{pmatrix}-(l-m+\frac{1}{2})^{1/2}Y_{l,m-1/2}\\(l+m+\frac{1}{2})^{1/2}Y_{l,m+1/2}\end{pmatrix}\end{aligned},
\end{equation}
thus one has $(\varphi_{\kappa,m},\varphi_{-\kappa,m})=(\chi_{j,m}^{(l)}, \chi_{j,m}^{(l')})$ with $l'=l+1$ for $j=l+1/2$ and $l'=l-1$ for $j=l-1/2$. One the other hand, the relation of $\kappa$, $l$ and $j$ are
\begin{equation}
    \kappa=\left\{\begin{array}{rl} -(l+1) & =-(j+1/2)~\quad {\rm for}\quad  j=l+1/2  \\
             l           & =+(j+1/2)~\quad {\rm for}\quad  j=l-1/2.
    \end{array}\right.
\end{equation}

The radial wave functions $F_{n,\kappa}(r)$ and $G_{n,\kappa}$ has been solved exactly as
\begin{eqnarray}\label{seq:r-function}
    F_{n,\kappa}(r)&=&+N\sqrt{m_c+E_{n,\kappa}}\cdot \rho^{s-1} e^{-\rho/2} \left[\left(\frac{(n'+s)m_c}{E_{n,\kappa}}-\kappa\right)F(-n',2s+1;\rho)-n' F(1-n',2s+1;\rho)\right]\nonumber\\
    G_{n,\kappa}(r)&=&-N\sqrt{m_c-E_{n,\kappa}}\cdot \rho^{s-1} e^{-\rho/2} \left[\left(\frac{(n'+s)m_c}{E_{n,\kappa}}-\kappa\right)F(-n',2s+1;\rho)+n' F(1-n',2s+1;\rho)\right],
\end{eqnarray}
where $F(a,b;z)$ is the confluent hypergeometric function, $N$ is a normalization factor makeing $\int\limits_0^\infty (G^2+F^2)r^2 dr =1$, and the parameters $s,n',n,\rho$ are defined by
\begin{eqnarray}\label{seq:params}
    s&=&\sqrt{\kappa^2-\alpha_s^2}\equiv \sqrt{(j+1/2)^2-\alpha_s^2}\nonumber\\
    \rho&=& 2\lambda r,~~(\lambda=\sqrt{m_c^2-E_{n,\kappa}^2})\nonumber\\
    n'&=&\frac{\alpha_s E_{n,\kappa}}{\lambda}-s=0,1,2,\ldots,\nonumber\\
    n &=& n'+|\kappa|=n'+j+\frac{1}{2}.
\end{eqnarray}
Thus the eigen energy $E_{n,\kappa}$ is expressed as
\begin{equation}
    E_{n,\kappa}=m_c\left[1+\frac{\alpha_s^2}{(n-|\kappa|+s)^2} \right]^{-1/2}=m_c\left[1+\frac{\alpha_s^2}{(n'+\sqrt{\kappa^2-\alpha_s^2})^2} \right]^{-1/2}.
\end{equation}

For hydrogen-like atoms, one has $\alpha_s\to Z\alpha\ll 1$, $s\approx |\kappa|$, $E_{n,\kappa}\approx m_c$, such that $G_{n,\kappa}$ is the small component, and $F_{n,\kappa}$ is the large component that reproduces the radial wave function in the non-relativistic limit. However, for a charmonium state, $\alpha_s \gg \alpha$ makes the relativistic correction (the smaller component $F$) more pronounced and changes the behavior of the radial wave function, especially in the $r\sim 0$ region. The ratio of the overall factors of $G_{n,\kappa}$ and $F_{n,\kappa}$ is
\begin{equation}
    R(n',\kappa)=\frac{\sqrt{m_c-E_{n,\kappa}}}{\sqrt{m_c+E_{n,\kappa}}}\approx \frac{\alpha_s}{n-|\kappa|+s}\sim \mathcal{O}(10^{-1}).
\end{equation}
Now let us make discussions on individual charmonium state.
\begin{itemize}
    \item $\eta_c$ and $J/\psi$: They are ground states (ignoring the spin-spin interaction temporarily) corresponding to the state of $n=1$, $n'=0$, $\kappa=-1$. The major relativistic effect is reflected by the divergence of the radial wave functions in a small region around $r=0$, namely
          \begin{equation}
              (F_{1,-1}, G_{1,-1})\sim r^{s-1}\sim r^{-\alpha_s^2/2}~~(r\to 0).
          \end{equation}
    \item $\chi_{c0}$: This is a $J^{PC}=0^{++}$ state, so the eigenvalue of $\hat{j}$ takes only $j=1/2$. Since it is an excited state, it is principal
          quantum number is $n=2$ which implies $n'=1$ and $\kappa=1$. The asymptotic behaviors of $F(r)$ and $G(r)$ for $r\to 0$ are similar to those of the $1S$ states
          \begin{equation}
              (F_{2,1}, G_{2,1})\sim r^{s-1}\sim r^{-\alpha_s^2/2}~~(r\to 0).
          \end{equation}
          For $(n,n',\kappa)=(2,1,1)$ state of hydrogen atom, the small component is very tiny and negligible and the prefactor of $F(-n',2s+1;\rho)$ in Eq.~(\ref{seq:r-function}) is almost the unit, such that the constant term is cancelled out and $F(-1,3;\rho)-F(0,3;\rho)\propto \rho$ meets the expectation of the non-relativistic limit. For $\chi_{c0}$, the ratio $R_{n',\kappa}\approx \alpha_s/2$, while constant term in the bracket of $G_{2,1}$ is roughly two, while the linear term of $F_{2,1}(r)$ is roughly $\rho/3$. So the contribution from $G_{2,1}(r)$ plays even a dominant role and governs the small $r$ behavior.
\end{itemize}


In order to obtain a clearer picture of the wave functions of $h_c, \chi_{c1},\chi_{c2}$, we give the explicit expressions of the charm quark wave function $\phi_{ljm} (l=1)$ (here $l=1$ means the orbital angular momentum in the large component and manifests the orbital parity $(-)^l$) as follows:
\begin{eqnarray}
    \phi_{1\frac{3}{2}\frac{3}{2}}&=&F_{2,-2} Y_{11} \alpha_{c}+iG_{2,-2}\left[\sqrt{\frac{4}{5}}Y_{22}\beta_{c}-\sqrt{\frac{1}{5}}Y_{21}\alpha_{c}\right]\nonumber\\
    \phi_{1\frac{3}{2}\frac{1}{2}}&=& F_{2,-2}\left[\sqrt{\frac{2}{3}} Y_{10}\alpha_c+\frac{1}{\sqrt{3}} Y_{11} \beta_c\right]\nonumber+iG_{2,-2}\left[\sqrt{\frac{3}{5}}Y_{21}\beta_c-\sqrt{\frac{2}{5}}Y_{20}\alpha_c\right]\\
    \phi_{1\frac{3}{2}-\frac{1}{2}} &=& F_{2,-2}\left[\frac{1}{\sqrt{3}}Y_{1-1}\alpha_c+\sqrt{\frac{2}{3}} Y_{10} \beta_c\right]\nonumber+iG_{2,-2}\left[\sqrt{\frac{2}{5}}Y_{20}\beta_c-\sqrt{\frac{3}{5}}Y_{2-1}\alpha_c\right]\\
    \phi_{1\frac{3}{2}-\frac{3}{2}}&=&F_{2,-2}Y_{1-1} \beta_c+iG_{2,-2}\left[\sqrt{\frac{1}{5}}Y_{2-1}\beta_{c}-\sqrt{\frac{4}{5}}Y_{2-2}\alpha_{c}\right]\nonumber\\
    \phi_{1\frac{1}{2}\frac{1}{2}}&=&F_{2,1}\left[-\frac{1}{\sqrt{3}}Y_{10}\alpha_c+\sqrt{\frac{2}{3}}Y_{11}\beta_c\right]\nonumber+iG_{2,1}Y_{00}\alpha_c\\
    \phi_{1\frac{1}{2}-\frac{1}{2}}&=&F_{2,1}\left[-\sqrt{\frac{2}{3}} Y_{1-1}\alpha_c+\frac{1}{\sqrt{3}} Y_{10}\beta_c\right]+iG_{2,1}Y_{00}\beta_c.
\end{eqnarray}
where $\alpha_{c}$ and $\beta_{c}$ are the spin-up and spin-down states of the charm quark, respectively, and the the radial coordinates $r$ of $F_{n,\kappa}(r)$ and $G_{n,\kappa}$ are omitted for convenience. Note that for $l=1$ states, the orbital angular momentum of the large component (proportional to $F_{n,\kappa}$) is $l=1$, while the small component (proportional to $G_{n,\kappa}$) has $l'=0$ or $l'=2$ which are the results of the relativistic effects since the orbital angular momentum is not conserved now.

Now we take into account the spin states of the anti-quark that provides the gluonic field for the charm quark. Let $\alpha_{\bar{c}}$ and $\beta_{\bar{c}}$ be spin-up and spin-down states of the charm anti-quark, then the total angular momentum $\hat{J}$ of a charmonium should be $\hat{J}=\hat{j}+\bar{s}_{c}$. We introduce the total spin wave functions
\begin{eqnarray}
    S_{00}&=&\frac{1}{\sqrt{2}}\left(\alpha_c\beta_{\bar{c}}-\beta_c\alpha_{\bar{c}}\right)\nonumber\\
    S_{10}&=&\frac{1}{\sqrt{2}}\left(\alpha_c\beta_{\bar{c}}+\beta_c\alpha_{\bar{c}}\right)\nonumber\\
    S_{11}&=&\alpha_c\alpha_{\bar{c}}\nonumber\\
    S_{1-1}&=&\beta_c\beta_{\bar{c}}.
\end{eqnarray}
Then the toal wave function of the scalar charmonium $\chi_{c0}(1^3P_0)$ is
\begin{equation}
    \begin{aligned}
        \Psi_{00}(0^{++})=\frac{F_{2,1}}{\sqrt{3}}\left(Y_{11}S_{1-1}+Y_{1,-1}S_{11}-Y_{10}S_{10}\right).
    \end{aligned}
\end{equation}
One can easily show $|\Psi_{00}(0^{++})|^2\propto 1/(4\pi)$ is a constant.

The $l=1$ state of the charm quark can make $J^P=1^+$ states with the anti charm quark through the $j=1/2$ and $j=3/2$ states, whose wave functions are
\begin{equation}
    \begin{aligned}
         & \Psi_{11}(1^{+})=\frac{1}{\sqrt{2}}\left[\frac{\sqrt{3}}{2}\phi_{1\frac{3}{2}\frac{3}{2}}\beta_{\bar{c}}-\frac{1}{2}\phi_{1\frac{3}{2}\frac{1}{2}}\alpha_{\bar{c}}+\phi_{1\frac{1}{2}\frac{1}{2}}\alpha_{\bar{c}}\right]=C_-\Psi_{11}(1^{+-})+C_+ \Psi_{11}(1^{++})             \\
         & \Psi_{10}(1^{+})=\frac{1}{2}\left[\phi_{1\frac{3}{2}\frac{1}{2}}\beta_{\bar{c}}-\phi_{1\frac{3}{2}-\frac{1}{2}}\alpha_{\bar{c}}+\phi_{1\frac{1}{2}\frac{1}{2}}\beta_{\bar{c}}+\phi_{1\frac{1}{2}-\frac{1}{2}}\alpha_{\bar{c}}\right]=C_-\Psi_{10}(1^{+-})+C_+ \Psi_{10}(1^{++}) \\
         & \Psi_{1-1}(1^{+})=\frac{1}{\sqrt{2}}\left[\frac{-\sqrt{3}}{2}\phi_{1\frac{3}{2}-\frac{3}{2}}\alpha_{\bar{c}}+\frac{1}{2}\phi_{1\frac{3}{2}-\frac{1}{2}}\beta_{\bar{c}}+\phi_{1\frac{1}{2}-\frac{1}{2}}\beta_{\bar{c}}\right]C_-\Psi_{1-1}(1^{+-})+C_+ \Psi_{1-1}(1^{++})
    \end{aligned}
\end{equation}
where $\Psi_{1M}(1^{+-})$ and $\Psi_{1M}(1^{++})$ are the wave function of $h_c$ ($S=0$) and $\chi_{c1}$ ($S=1$), and $C_\pm$ are combination coefficients. The explicit expressions of $\Psi_{1M}(1^{+\mp})$ are
\begin{eqnarray}
    \Psi_{1M}(1^{+-}) &=& \left(\frac{1}{\sqrt{3}}F_{2,1}-\sqrt{\frac{2}{3}}F_{2,-2}\right)Y_{1M}S_{00}\nonumber\\
    \Psi_{11}(1^{++}) &=& \left(\frac{1}{\sqrt{6}}F_{2,-2}+\frac{1}{\sqrt{3}}F_{2,1}\right)\left(Y_{11}S_{10}-Y_{10}S_{11}\right)+i\left(G_{2,1}Y_{00}S_{11}-G_{2,-2}\mathcal{D}_{1}\right)\nonumber\\
    \Psi_{10}(1^{++}) &=& \left(\frac{1}{\sqrt{6}}F_{2,-2}+\frac{1}{\sqrt{3}}F_{2,1}\right)\left(Y_{11}S_{1-1}-Y_{1-1}S_{11}\right)+i\left(G_{2,1}Y_{00}S_{10}-G_{2,-2}\mathcal{D}_{0}\right)\nonumber\\
    \Psi_{1-1}(1^{++}) &=& \left(\frac{1}{\sqrt{6}}F_{2,-2}+\frac{1}{\sqrt{3}}F_{2,1}\right)\left(Y_{10}S_{1-1}-Y_{1-1}S_{10}\right)+i\left(G_{2,1}Y_{00}S_{1-1}-G_{2,-2}\mathcal{D}_{-1}\right),
\end{eqnarray}
where
\begin{equation}
    \begin{aligned}
        \mathcal{D}_{+1} & =\sqrt{\frac{3}{5}}Y_{22}S_{1,-1}-\sqrt{\frac{3}{10}}Y_{21}S_{10}+\sqrt{\frac{1}{10}}Y_{20}S_{11}, \\\mathcal{D}_{0}&=\sqrt{\frac{3}{10}}Y_{21}S_{1,-1}-\sqrt{\frac{2}{5}}Y_{20}S_{10}+\sqrt{\frac{3}{10}}Y_{2,-1}S_{11},\\\mathcal{D}_{-1}&=\sqrt{\frac{1}{10}}Y_{20}S_{1,-1}-\sqrt{\frac{3}{10}}Y_{2,-1}S_{10}+\sqrt{\frac{3}{5}}Y_{2,-2}S_{11}.
    \end{aligned}
\end{equation}
Similarly we have wave functions for the $2^{++}$ state $\chi_{c2}$:
\begin{equation}\label{seq:2pp-wave}
    \begin{aligned}
        \Psi_{22}(2^{++})  & =\phi_{1\frac{3}{2}\frac{3}{2}}\alpha_{\bar{c}}                                                                                                                                            \\
                           & =F_{2,-2}Y_{11}S_{11}+iG_{2,-2}\left[\sqrt{\frac{2}{3}}Y_{22}S_{10}-\sqrt{\frac{1}{3}}Y_{21}S_{11}\right]                                                                                  \\
        \Psi_{21}(2^{++})  & =\frac{1}{2}\left[\sqrt{3}\phi_{1\frac{3}{2}\frac{1}{2}}\alpha_{\bar{c}}+\phi_{1\frac{3}{2}\frac{3}{2}}\beta_{\bar{c}}\right]                                                              \\
                           & =\frac{1}{\sqrt{2}}F_{2,-2}\left[Y_{10}S_{11}+Y_{11}S_{10}\right]+iG_{2,-2}\left[\sqrt{\frac{1}{3}}Y_{22}S_{1-1}+\sqrt{\frac{1}{6}}Y_{21}S_{10}-\sqrt{\frac{1}{2}}Y_{20}S_{11}\right]      \\
        \Psi_{20}(2^{++})  & =\frac{1}{\sqrt{2}}\left[\phi_{1\frac{3}{2}-\frac{1}{2}}\alpha_{\bar{c}}+\phi_{1\frac{3}{2}\frac{1}{2}}\beta_{\bar{c}}\right]                                                              \\
                           & =F_{2,-2}\left[\frac{Y_{11}S_{1,-1}+2Y_{10}S_{10}+Y_{1,-1}S_{11}}{\sqrt{6}}\right]+iG_{2,-2}\left[-\sqrt{\frac{1}{2}}Y_{21}S_{1,-1}+\sqrt{\frac{1}{2}}Y_{2,-1}S_{11}\right]                \\
        \Psi_{2-1}(2^{++}) & =\frac{1}{2}\left[\phi_{1\frac{3}{2}-\frac{3}{2}}\alpha_{\bar{c}}+\sqrt{3}\phi_{1\frac{3}{2}-\frac{1}{2}}\beta_{\bar{c}}\right]                                                            \\
                           & =\frac{1}{\sqrt{2}}F_{2,-2}\left[Y_{10}S_{1-1}+Y_{1-1}S_{10}\right]+iG_{2,-2}\left[-\sqrt{\frac{1}{3}}Y_{2-2}S_{11}-\sqrt{\frac{1}{6}}Y_{2-1}S_{10}+\sqrt{\frac{1}{2}}Y_{20}S_{1-1}\right] \\
        \Psi_{2-2}(2^{++}) & =\phi_{1\frac{3}{2}-\frac{3}{2}}\beta_{\bar{c}}                                                                                                                                            \\
                           & =F_{2,-2}Y_{1-1}S_{1-1}+iG_{2,-2}\left[-\sqrt{\frac{2}{3}}Y_{2-2}S_{10}+\sqrt{\frac{1}{3}}Y_{2-1}S_{1-1}\right]                                                                            \\
    \end{aligned}
\end{equation}

The above wave functions give us the following implications:
\begin{itemize}
    \item $h_c$: It has $j=1/2$ ($\kappa=1$) and $j=3/2$ ($\kappa=-2$) components and its principal quantum number is $n=2$. The small $r$ behaviors of $F_{2,1}(r)$ and $F_{2,-2}$ are
          \begin{equation}
              F_{2,1}(r) \sim r^{-\alpha_s^2/2},\quad F_{2,-2}(r)\sim r^{1-\alpha_s^2/8}\quad (r\to 0),
          \end{equation}
          since $s=\sqrt{\kappa^2-\alpha_s^2}$.
          So $F_{2,-2}(r)$ approaches to zero when $r\to 0$, while $F_{2,1}(r=0)$ deviates from zero and also has a weak divergence at $r=0$.
    \item $\chi_{c1}$: The large component has the similar radial behavior to that of $h_c$. In addition, the relative effects are manifested by the appearance of $l=0$ and $l=2$ partial wave in the small component. The angular distribution due to the $l=2$ component is observed in the $|\chi_{\chi_{c1}}(\theta,\phi)|^2$, as shown the bottom left panel in Fig.`\ref{sfig:3-d}.
    \item $\chi_{c2}$: This is a $J^{PC}=2^{++}$ state, so the eigenvalue of $\hat{j}$ takes only $j=3/2$ ($\kappa=-2$). The principal quantum number is $n=2$ which implies $n'=0$. The asymptotic behaviors of $F_{2,-2}(r)$ and $G_{2,-2}(r)$ for $r\to 0$ are
          \begin{equation}
              (F_{2,-2}, G_{2,-2})\sim r^{s-1}\sim r^{1-\alpha_s^2/8}~~(r\to 0).
          \end{equation}
          So the radial wave function $\phi_{\chi_{c2}}(r)$ has a similar $r$-behavior to that of the $1P$ state in the nonrelativistic limit. On the other hand the lower component that is proportional to $G_{2,-2}(r)$ is also important. One can see from Eq.~(\ref{seq:2pp-wave}) that the lower component has $l=2$ partial wave, which can be visualized by its angular distribution in Fig.~\ref{sfig:3-d}, where a quadripole distribtuion is superposed on the $P$-wave distribution.
\end{itemize}

\begin{figure*}[t]
\includegraphics[width=0.23\linewidth]{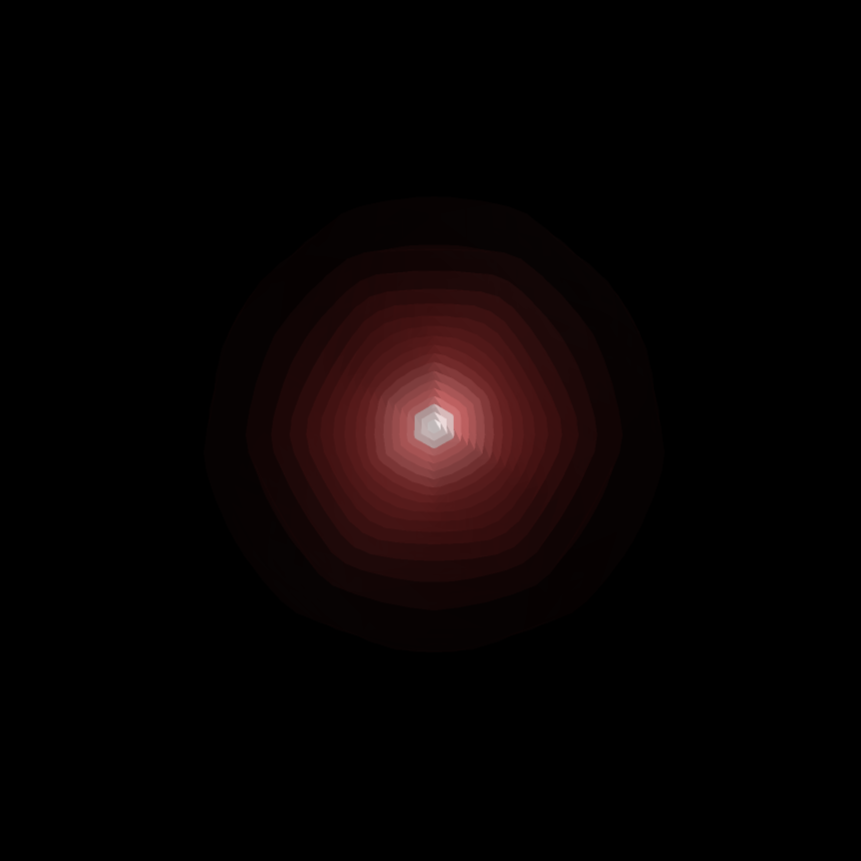}
\includegraphics[width=0.23\linewidth]{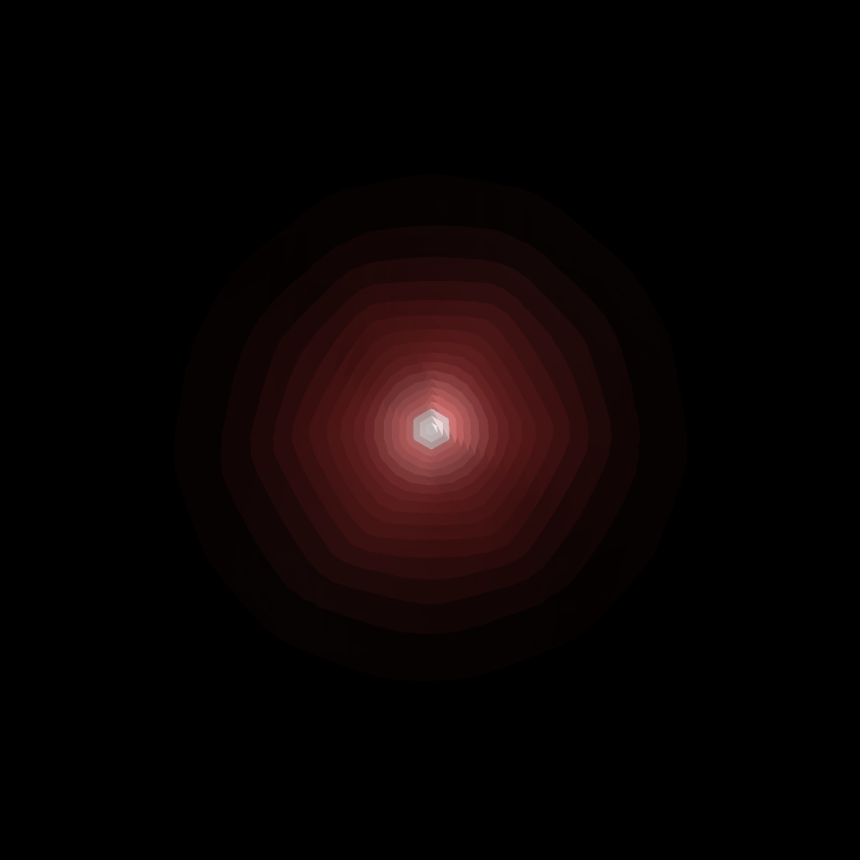}
\includegraphics[width=0.23\linewidth]{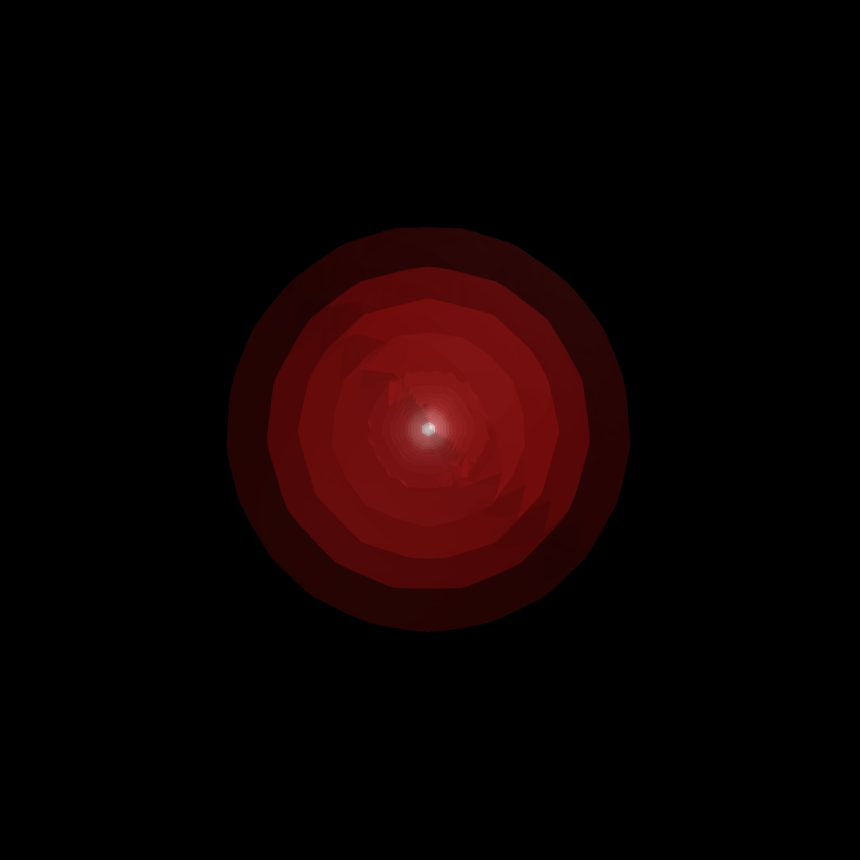}
\includegraphics[width=0.23\linewidth]{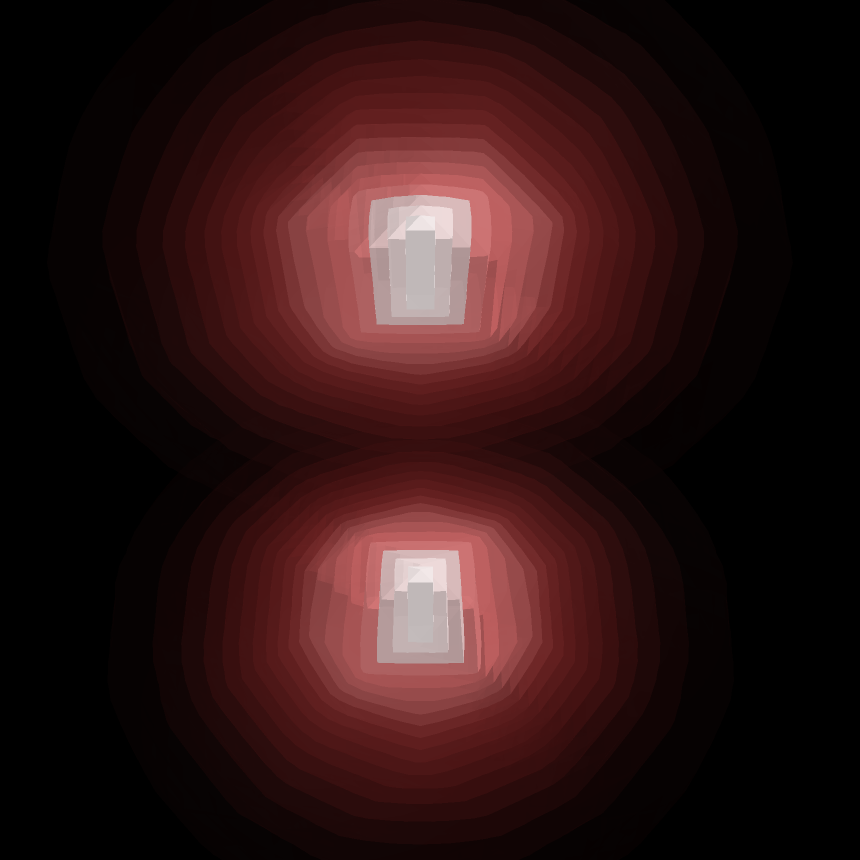}\\
\includegraphics[width=0.23\linewidth]{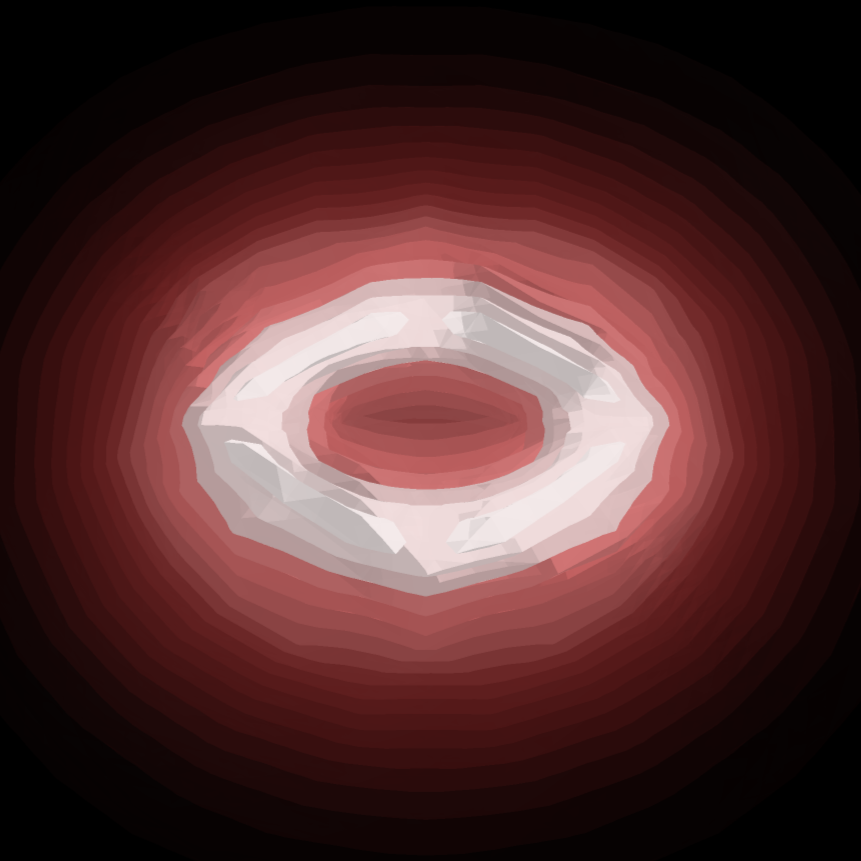}
\includegraphics[width=0.23\linewidth]{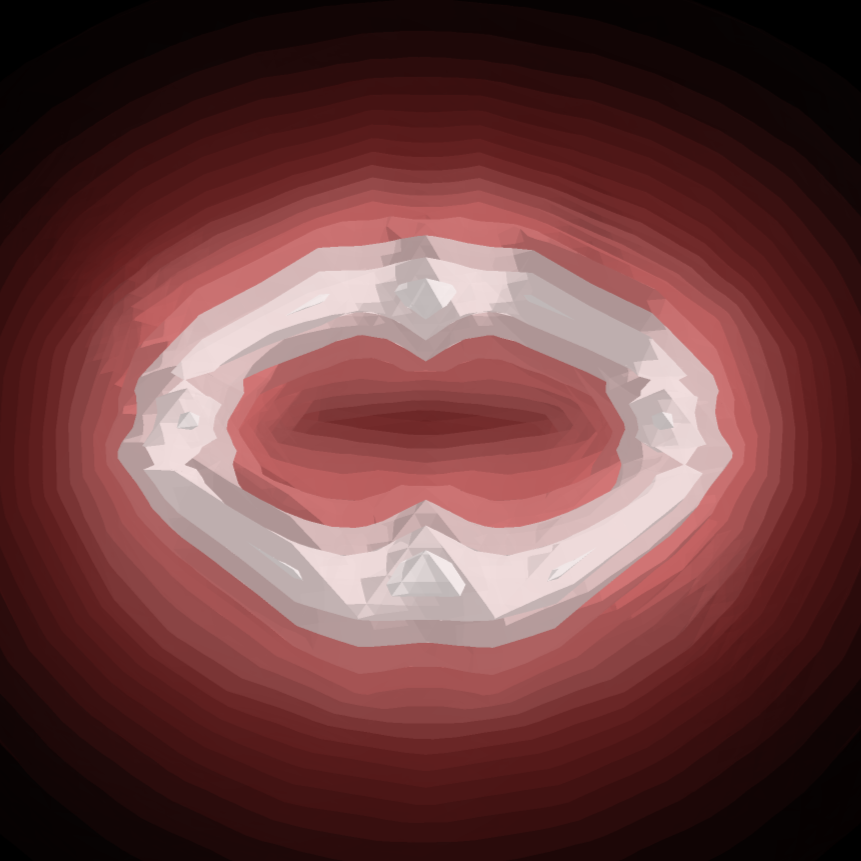}
\includegraphics[width=0.23\linewidth]{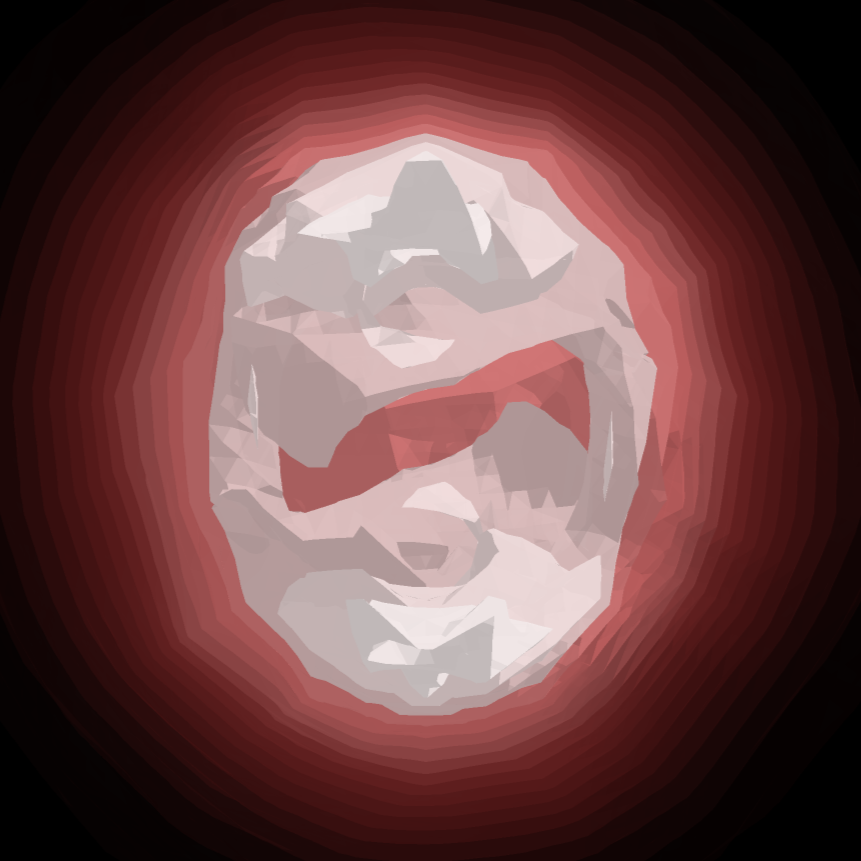}
\includegraphics[width=0.23\linewidth]{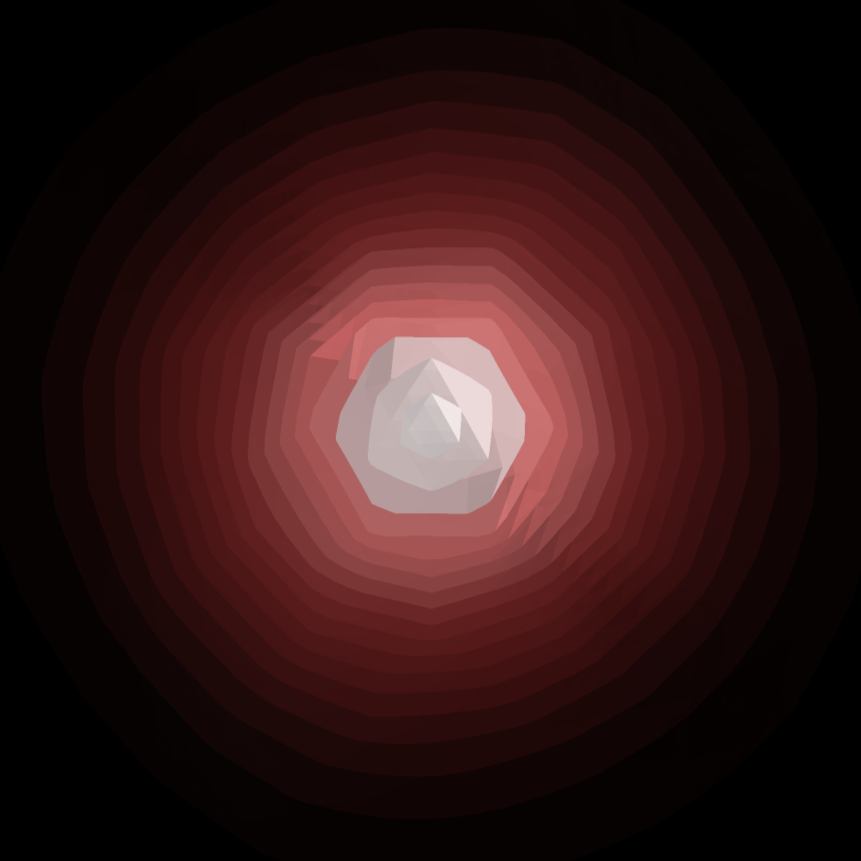}
\caption{\justifying
    The angular distributions of $1S$, $1P$ charmonia and $0^{+-}$ and $1^{-+}$ charmoniumlike states based on the raw data from lattice QCD calculation. In the top row, the panels from left to right are the angular distributions of $\eta_c$, $J/\psi$, $\chi_{c0}$, $h_c$, respectively. In the bottom row, the panels are arranged from left to right for $\chi_{c1}$, $\chi_{c2}$, $0^{+-}$ and $1^{-+}$ states.\label{sfig:3-d}}
Three-dimensional probability-density renderings $\rho(\vec{r})=|\psi(\vec{r})|^{2}$ for six low-lying charmonium states in their rest frame, displayed left-to-right as $\eta_{c}(0^{-+}), J/\psi(1^{--}), \chi_{c0}(0^{++}), h_{c}(1^{+-}), \chi_{c1}(1^{++})$, and $\chi_{c2}(2^{++})$.All states are shown with their spin (polarization) axis aligned along the $z$‑direction. Warmer (yellow) tones correspond to higher $\rho$, cooler (dark red-black) tones to lower $\rho$. All panels share a common logarithmic colour scale and cover the same spatial volume, making nodal structures and radial tails directly comparable across states.\label{sfig:3-d}
\end{figure*}


In the $J=2$ channel we isolate the quadrupole polarization along the lattice $z$-axis by projecting our wave function onto the three-dimensional $T_2$ irrep of the cubic group. Our graph then shows the spatial distribution of this $z$ -polarization. The complete unitary change of basis from the continuum spherical harmonics $\{\dot{Y_{2,m}}(r)\}_{m=-2}^2$ to the Cartesian linear-polarization components $\{ \psi _{xx- yy}$, $\psi _{xy}$, $\psi _{xz}$, $\psi _{yz}$, $\psi _{zz}\}$
is as follows:
$$\begin{aligned}\psi_{xx-yy}(r)&=\frac{Y_{2,2}(r)+Y_{2,-2}(r)}{\sqrt{2}},\\\psi_{xy}(r)&=\frac{Y_{2,2}(r)-Y_{2,-2}(r)}{i\sqrt{2}},\\\psi_{xz}(r)&=\frac{Y_{2,1}(r)+Y_{2,-1}(r)}{\sqrt{2}},\\\psi_{yz}(r)&=\frac{Y_{2,1}(r)-Y_{2,-1}(r)}{i\sqrt{2}},\\\psi_{zz}(r)&=Y_{2,0}(r)\:.\end{aligned}$$

Under the octahedral group $O_{h}$, this five-dimensional space decomposes as
$5\to E(2)\oplus T_2(3)$, where the two-dimensional $E$ irrep is spanned by $\{\psi_{xx-yy},\psi_{zz}\}$ and the three-dimensional $T_2$
irrep is spanned by $\{\psi_{xy},\psi_{xz},\psi_{yz}\}.$
By substituting the spherical-harmonic expansion of the wave function into our projection formula, one finds
$$\psi_{xy}(r) = \frac{1}{\sqrt{2}i}[F_{2,-2}(Y_{11}S_{11}-Y_{1-1}S_{1-1})+iG_{2,-2}(\sqrt{\frac{2}{3}}(Y_{22}-Y_{2-2})S_{10}-\sqrt{\frac{1}{3}}Y_{21}S_{11}-\sqrt{\frac{1}{3}}Y_{2,-1}S_{1,-1})]$$
This explicit form makes transparent how the quadrupole component is encoded in the resulting spatial distribution.

\subsection{S2. $1S-2S$ mass split in a Cornell potential}
A lattice QCD study~\cite{Ma:2019hsm} observes that, for a $1^{-+}$ charmoniumlike hybrid, the separation $r=|\vec{r}|$ between the color octet $c\bar{c}$ (in a ${}^3S_1$ state) component and the chromomagnetic (CM) excitation is a meaningful dynamical variable, and the BS wave functions with respect to $\rho$  exhibits the nodal behavior of a two-body $nS$ radial wave function. On the other hand, previous lattice calculations showed that the potential between two static color charges
has a Casimir scaling behavior~\cite{Bali:2000un}
\begin{equation}
    V^{(R)}(r)=V^{(R)}_0+\frac{C_2(R)}{C_2(F)}\left(-\frac{4\alpha_s}{3r}+\sigma r\right),
\end{equation}
where $C_2(R)$ is the Casimir eigenvalue of the irreducible representation (irreps) $R$ and $R^*$ of the static charges. For color octet (the adjoint irreps) one has $C_2(A)=3$ and for the fundamental irreps one has $C_2(F)=4/3$. If the hybrid can be viewed as a CM excitation moving in this Cornell-type potential generated by the color octet $c\bar{c}$, one can solve the Sch\"{o}dinger equation of this two-body following the procedure in Ref.\cite{Eichten:1978tg} to obtain the spectrum of this system. The $S$-wave ($l=0$) radial equation is written as
\begin{equation}
    -\frac{1}{2\mu}u^{\prime\prime}(r)+\left(-\frac{3\alpha_s}{r}+\frac{9}{4}\sigma r\right)u(r)=Eu(r),
\end{equation}
where $\mu$ is the reduced mass of the system. The eigen-energy $E_n$ of the $nS$ state is
\begin{equation}\label{seq:eigen}
    E_n=\left(\left(\frac{9\sigma}{4}\right)^2 \frac{1}{2\mu}\right)^{1/3} d_n,
\end{equation}
where the eigenvalues $d_n$ are derived numerically with given parameters. If we take the effective masses of the color octet $c\bar{c}$ block and the CM excitation to be 3.0 GeV and 0.8 GeV, respectively, and use the values $\alpha_{s}\approx0.3$, $\sigma\approx0.25\,\text{GeV}^{2}$, then we obtain
\begin{equation}
    d_1\approx1.556,\quad d_2\approx3.572.
\end{equation}
Putting these values into Eq.~(\ref{seq:eigen}) we get the $1S-2S$ mass splitting
\begin{equation}
    \Delta m(2S-1S)\approx1.272\mathrm{~GeV}.
\end{equation}
\subsection{S3. Fitting details}

We extract the effective masses from the two-point correlation functions in each channel: the individual markers show the lattice data, while the shaded band indicates the corresponding fit. A temporal separation of $\Delta t = 24$ was selected to maximise ground-state dominance and, at the same time, keep statistical uncertainties under control.
\begin{figure*}[htbp]
    \includegraphics[width=0.23\linewidth]{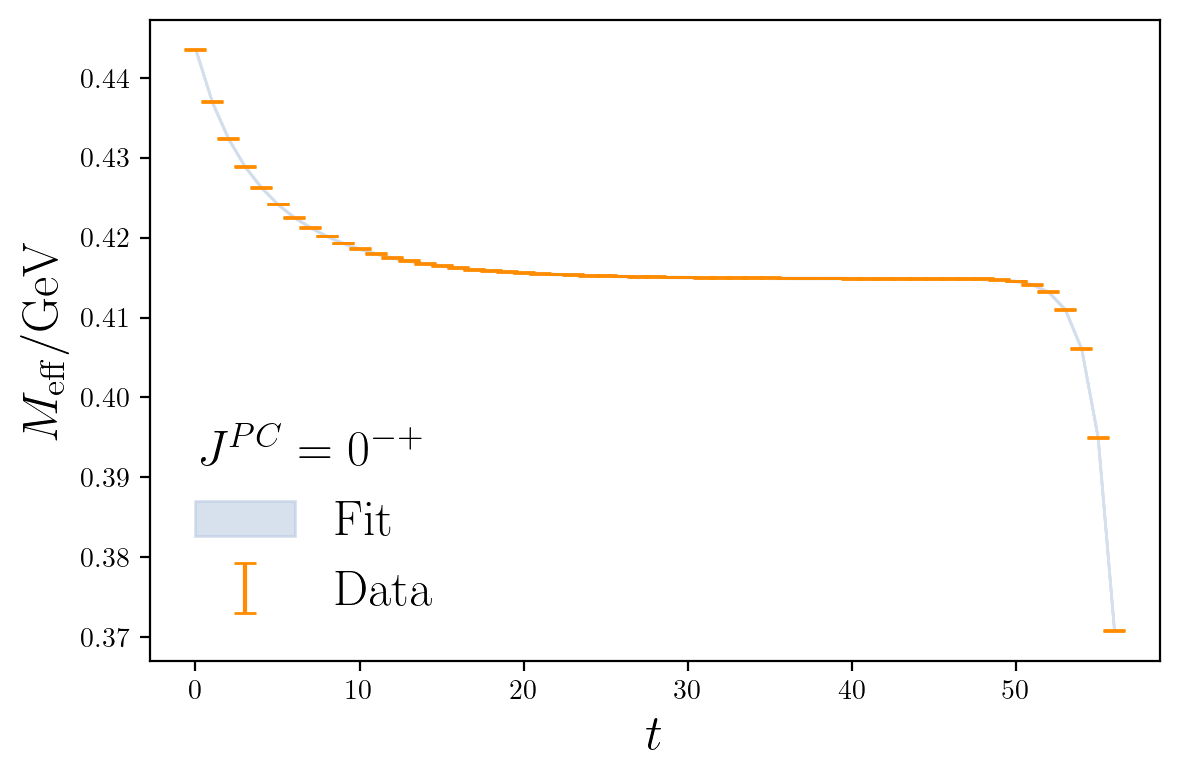}
    \includegraphics[width=0.23\linewidth]{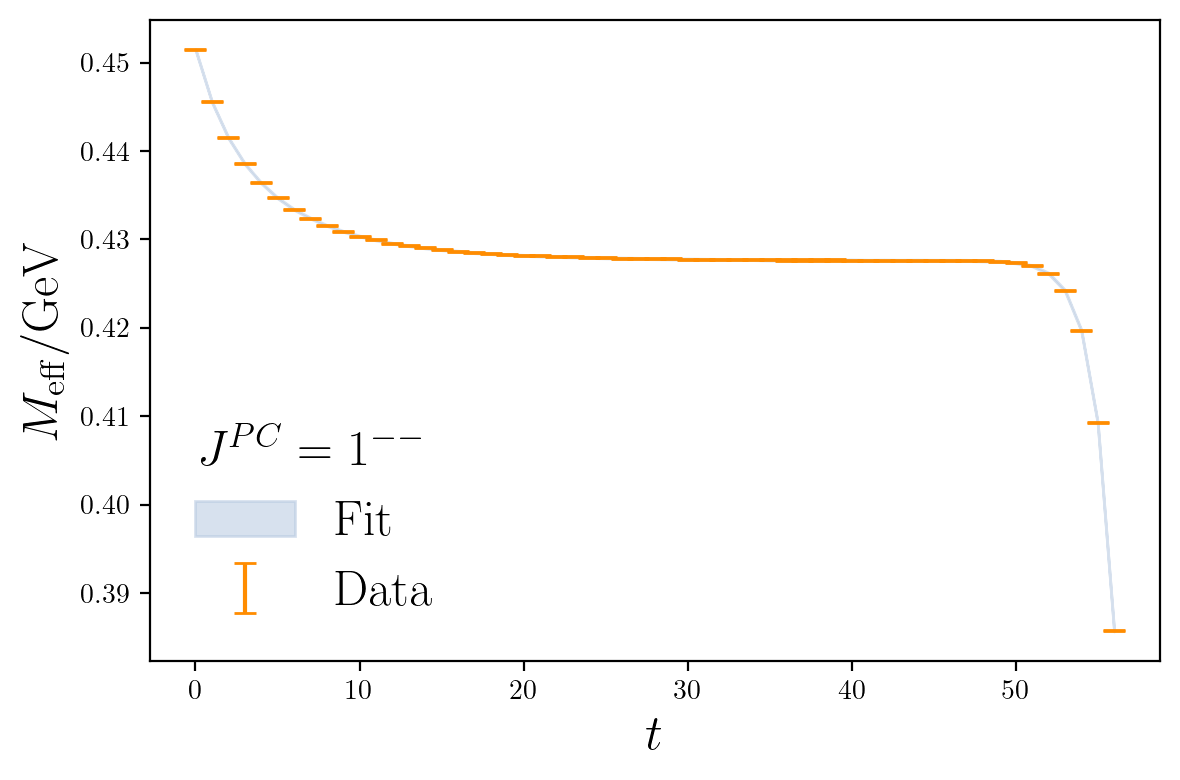}
    \includegraphics[width=0.23\linewidth]{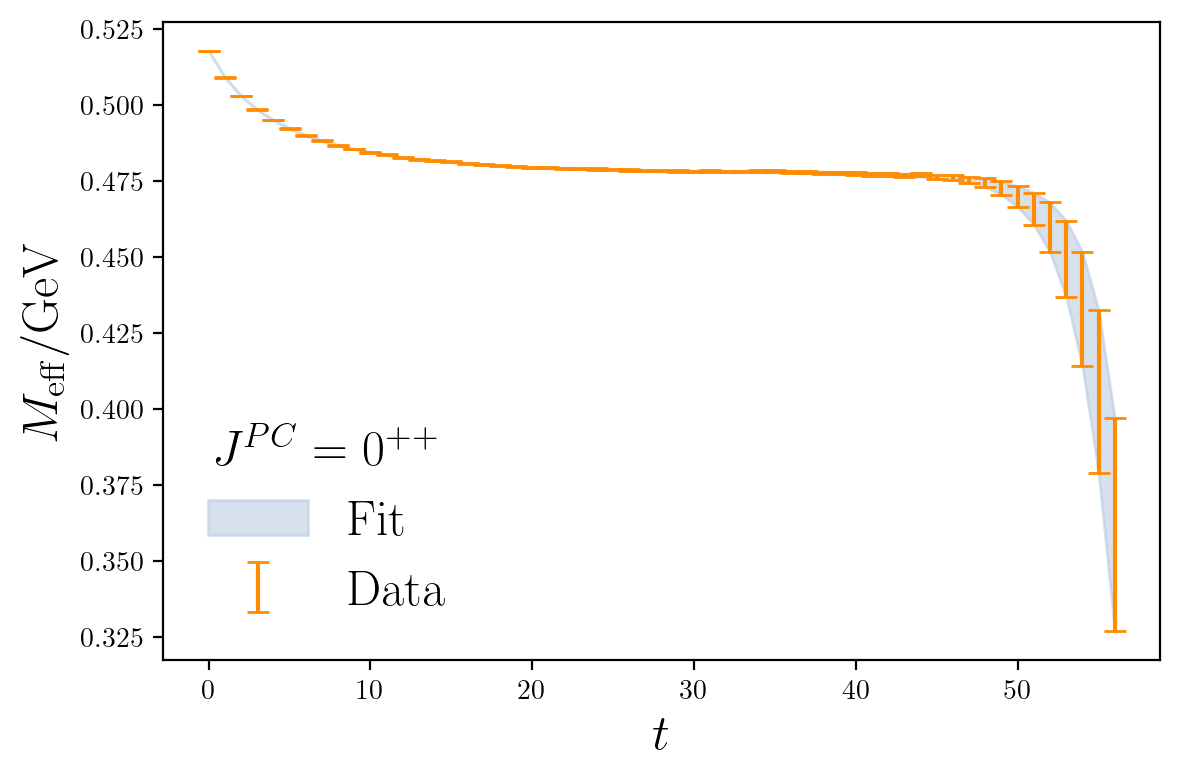}
    \includegraphics[width=0.23\linewidth]{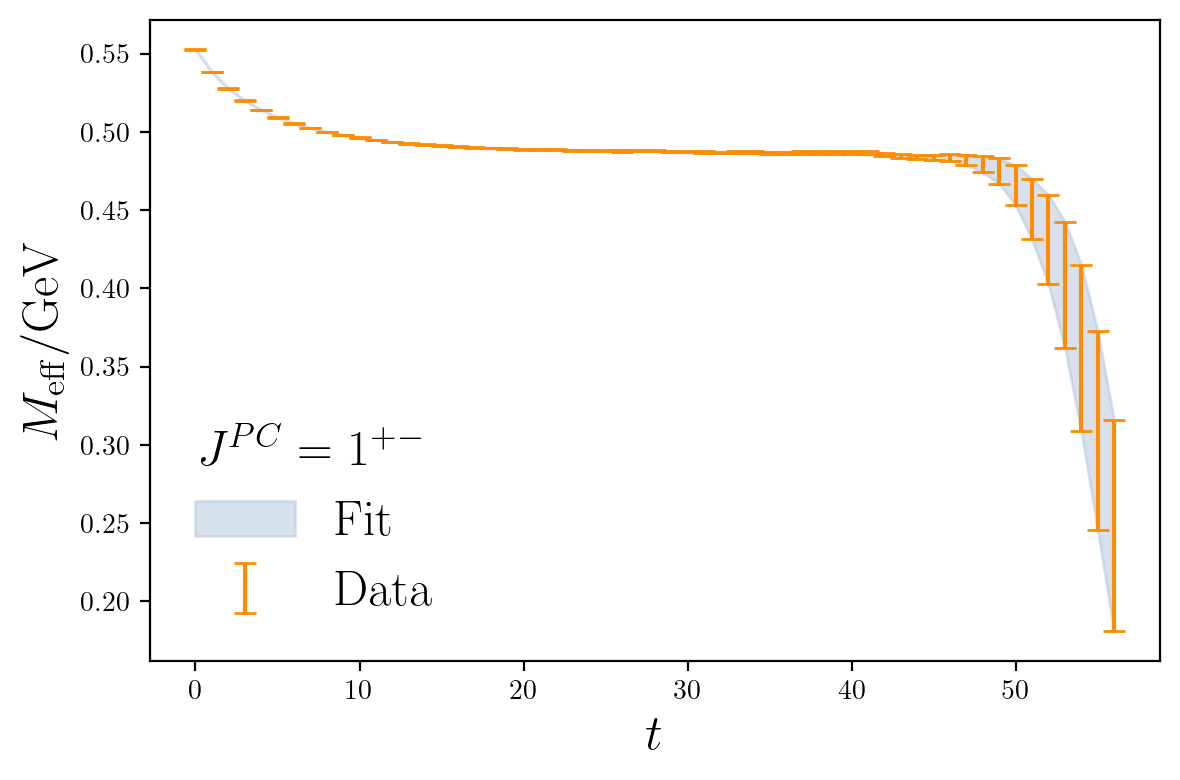}\\
    \includegraphics[width=0.23\linewidth]{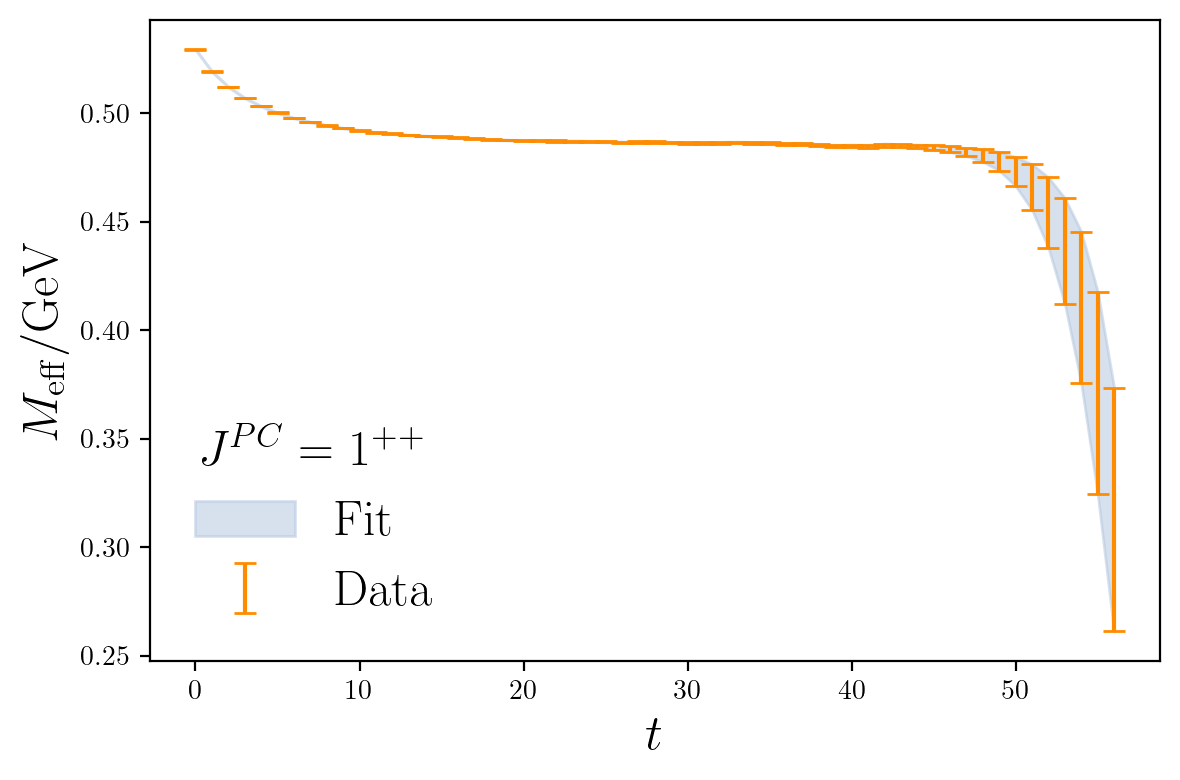}
    \includegraphics[width=0.23\linewidth]{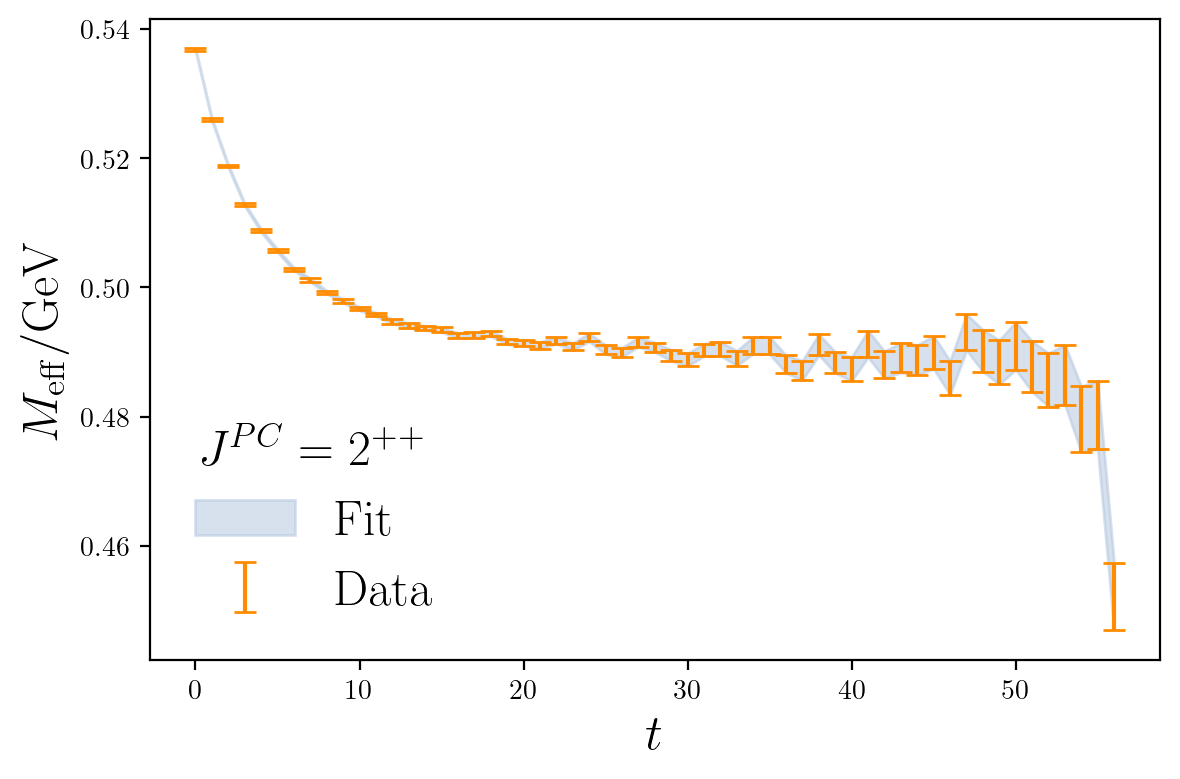}
    \includegraphics[width=0.23\linewidth]{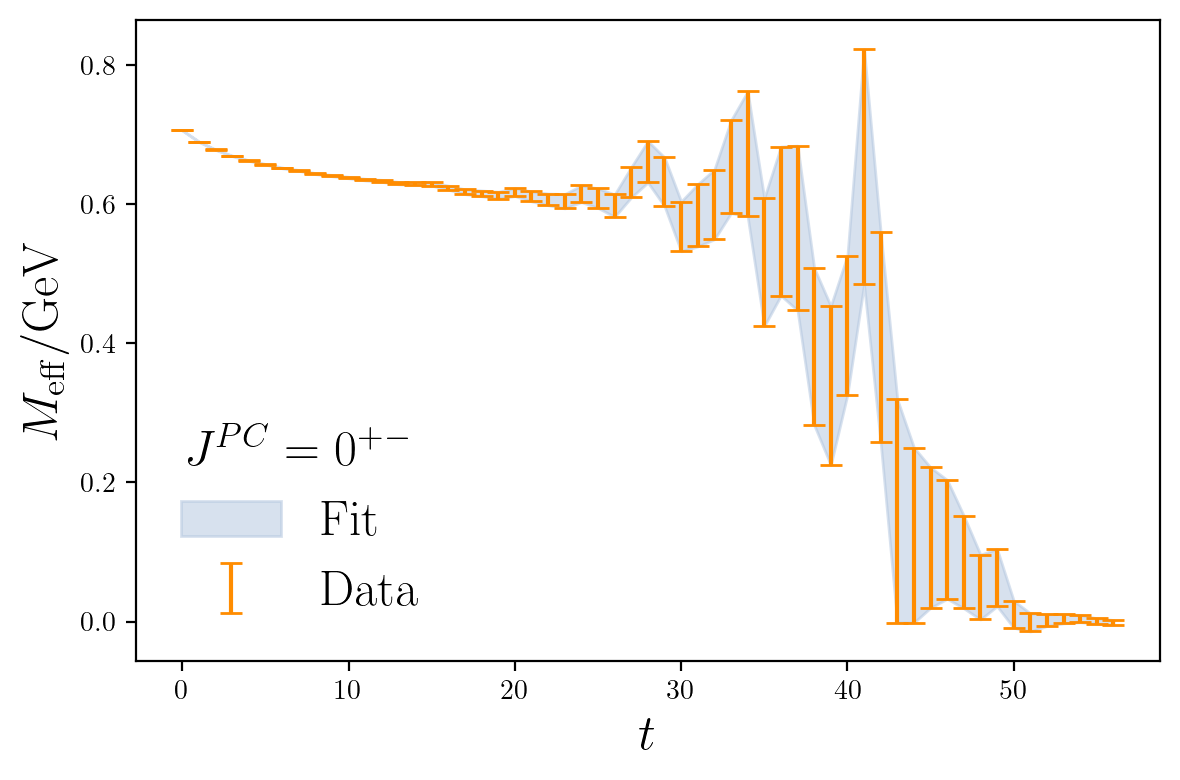}
    \caption{\justifying
        The effective masses of different charmonium-like state on L16. The data
        points are the lattice results, and the colored bands are the two-mass-term fits with the fitted parameters.}
\end{figure*}
\end{document}